\documentclass[aps,pre,twocolumn,showpacs]{revtex4}
\usepackage{amsfonts} \usepackage{amsmath} \usepackage{amsthm} \usepackage{graphics} \usepackage{graphicx} \usepackage{subfigure} \usepackage{amssymb}
\usepackage{graphicx} \usepackage{color}

\begin{document}

\title{Alfv\'en waves and ideal two-dimensional Galerkin truncated magnetohydrodynamics}
\author{Giorgio Krstulovic$^{1,2}$, Marc-Etienne Brachet$^1$ and Annick Pouquet$^3$}
\affiliation{$^1$Laboratoire de Physique Statistique de l'\'Ecole Normale Sup\'erieure, 
associ\'e au CNRS et aux Universit\'es ParisVI et VII, 24 Rue Lhomond, 75231 Paris, France. \\
 $^2$ Laboratoire Cassiop\'ee, Observatoire de la C\^ote dÕAzur, CNRS, Universit\'e de Nice Sophia-Antipolis, Bd. de l'Observatoire, 06300 Nice, France\\
 $^3$Computational and Information Systems Laboratory, National Center for Atmospheric Research, 
         P.O. Box 3000, Boulder, Colorado 80307-3000, USA.}
\date{\today}

\begin{abstract}
We investigate numerically the dynamics of two-dimensional Euler and ideal magnetohydrodynamics (MHD) flows in systems with a finite number of modes, up to $4096^2$,  for which several quadratic invariants are preserved by the truncation and the statistical equilibria are known. Initial conditions are the Orszag-Tang vortex with a neutral X-point centered on a stagnation point of the velocity field in the large scales.  In MHD, we observe that the total energy spectra at intermediate times and intermediate scales correspond to the interactions of eddies and waves, $E_T(k)\sim k^{-3/2}$. 
Moreover, no dissipative range is visible neither for Euler nor for MHD in two dimensions; in the former case, this may be linked to the existence of a vanishing turbulent viscosity whereas in MHD, the numerical resolution employed may be insufficient. When imposing a uniform magnetic field to the flow, we observe a lack of saturation of the formation of small scales together with a significant slowing-down of their equilibration, 
with however a cut-off independent partial thermalization being reached at intermediate scales.
\end{abstract}
\maketitle

\section{Introduction}\label{Sec:Intro}

A theory of turbulent flows is still eluding us, and yet such flows are ubiquitous in nature and are an integral part of the problem of weather prediction or of  climate assessment, as well as for the formation and prediction of extreme events such as tornadoes and hurricanes; they are also an important element in the understanding of the dynamics of planets and the heliosphere, stars and beyond. What is lacking is a statistical description of the small scales, and a prediction of long-time large-scale dynamics with ensuing modified transport properties. Of course, since Onsager (1949) \cite{onsager} who considered an ensemble of point vortices in two space dimensions (2D), and T.D. Lee (1952) \cite{LEE:1952p4100} who pioneered the study of the behavior of a truncated system of modes both for inviscid fluids and ideal magnetohydrodynamics (MHD) in three dimensions (3D), we know that equipartition of energy among the modes obtain in the simplest (3D, non helical) case, whereas spectra peaking at the gravest mode in the 2D case are possible; these latter solutions can be viewed as precursors of inverse cascades toward large scales, observed in 2D for forced dissipative fluids and for MHD \cite{Kraichnan:1967p3033,matthaeus_mhd} (see \cite{Kraichnan:1980p248} for a review), as well as in 3D MHD \cite{mazure} (see \cite{pouquet_houch} for a review). Numerous extensions of the 2D problems introducing potential vorticity lead to the same prediction of an inverse cascade of energy, as in the quasi-geostrophic case.

It has been argued that inviscid dynamics is in fact a good indicator of the behavior of turbulent flows in the presence of forcing and dissipation. This is certainly the case for scales larger than the forcing scale $\ell_F=2\pi/k_F$ when an inverse cascade is present, but evidence for scales smaller than $\ell_F$ has been lacking until recently. It was shown, using high resolution ideal runs in 3D for Euler flows that at intermediate times and intermediate scales, a Kolmogorov energy spectrum (hereafter, K41), $E(k)\sim k^{-5/3}$, is observed, as well as a dissipative range at the end of the K41 spectrum; the latter range can be attributed to an eddy viscosity due to the thermalized high-k modes, $E(k)\sim k^2$ \cite{Cichowlas:2005p1852} (see also \cite{Krstulovic:2009p4422} in the helical case when velocity-vorticity correlations are included, and \cite{mininni1} in the 3D rotating case, with or without helicity).
One of the most striking result on the truncated Euler equations is thus that the thermalized zone progressively extends to large scales finally covering the whole spectrum as expected, but that the intermediate scales are found to follow an effective Navier-Stoke equation \cite{Krstulovic:2008p428,Krstulovic:2009p4422} even though dissipation is absent.

Two-dimensional MHD is special for at least two reasons. 
First, the 2D ideal case possesses an infinite number of invariants, even though only the quadratic ones are being preserved in the truncated system. Nevertheless, the afore-mentioned predictions of inverse cascades have been verified in many instances \cite{Kraichnan:1980p248}: one can indeed argue that other invariants, such as any power of the vorticity field integrated over space, will be dissipated efficiently in the realistic case in which forcing and viscosity are included.
Furthermore, in MHD, there is a controversy as to what is the energy spectrum in the forced dissipative case: it could be a classical Kolmogorov spectrum, either isotropic or anisotropic, in that latter case with a dependence on $k_{\perp}$ (referring to the direction perpendicular to an imposed strong uniform magnetic field ${\bf b}_0$, of magnitude $b_0$ \cite{Goldreich:1995p3870}). Or it could be an Iroshnikov-Kraichnan spectrum (hereafter, IK \cite{Iroshnikov:1964p3299, Kraichnan:1965p2693}) stemming from the interactions of Alfv\'en waves and turbulent eddies and leading to the slowing-down of the nonlinear cascade when waves are strong; in fact, this latter solution is compatible with the prediction of weak turbulence theory for MHD \cite{Galtier:2000p3795} when the resulting spectrum is isotropized. Moreover, it has been argued by several authors that there is no complete universality in MHD, including in the absence of forcing and with ${\bf b}_0 \equiv 0$ (see \cite{lee2} and references therein).

In this paper, we thus investigate the properties of ideal two-dimensional MHD and Euler flows.
The next section explains the procedure we follow and recalls several known properties of ideal 2D MHD; in \S \ref{s:b0=0}, we give results in the absence of imposed magnetic field, ${\bf b}_0\equiv 0$. In \S \ref{s:b0not=0}, we analyze the slowing-down of the dynamics for ${\bf b}_0\not= 0$; finally, \S \ref{s:struct} briefly examine the structures that develop in space, and \S \ref{s:conclu} is the conclusion. The special (${\bf b}=0$) Euler case in two dimensions is treated in the Appendix.

\section{The procedure} \label{s:proc}

We begin by writing the ideal MHD equations in the incompressible case ($\nabla \cdot {\bf u}=0$, where ${\bf u}$ is the velocity field) by
introducing a pseudoscalar potential $\psi(x,y,t)$, the stream function, and the (scalar) magnetic potential $a(x,y,t)$, with ${\bf u}=\nabla \times \psi$ and ${\bf b}=\nabla \times a$, ${\bf b}$ being the magnetic induction, also divergence-free ($\nabla \cdot {\bf b}=0$) in the absence of magnetic monopoles:

\begin{eqnarray}
\frac{\partial \psi}{\partial t}&=&\frac{1}{\nabla^2}\{\psi,\nabla^2\psi\}-\frac{1}{\nabla^2}\{a,\nabla^2a\} \label{Eq:MHDPsipot}\\
\frac{\partial a}{\partial t}&=&\{\psi,a\} \ , \label{Eq:MHDapot}
\end{eqnarray}
where $\{f,g\}=\partial_xf\partial_yg-\partial_xg\partial_yf$ is the usual Poisson bracket. 
Regrouping some terms in equation (\ref{Eq:MHDapot}), it can be shown that the magnetic potential also satisfies:
\begin{equation}
\frac{\partial a}{\partial t}+{\bf u}\cdot{\bf \nabla}a=0
\end{equation}
and therefore $a$ is advected as a passive scalar by the fluid, even though the Lorentz force ${\bf j} \times {\bf b}$ acts on the fluid and breaks the conservation of vorticity in 2D, (with ${\bf j}=\nabla \times {\bf b}$ the current density).
Equations (\ref{Eq:MHDPsipot}, \ref{Eq:MHDapot}) conserve the total energy
\begin{eqnarray}
E&=&\frac{1}{2}\int d^2x\,[|{\bf u}|^2+|{\bf b}|^2] \ ;\\
\end{eqnarray}
they also have an infinite number of conserved quantities, the Casimirs, of the form
\begin{eqnarray}
\mathcal{C}&=&\int d^2x\,[f(a)+\nabla^2\psi g(a)] , \label{Eq:CasimirMHD}
\end{eqnarray}
where $f$ and $g$ are arbitrary functions.
Among them, two remarkable invariants are obtained for $f(a)=a^2, g(a)=0$ and for $f(a)=0,g(a)=-a$; the former one, denoted $A\equiv <a^2>$, is the conserved square magnetic potential and the latter is  called the cross helicity which can also be written as 
$H_{c}=\int d^2x\, {\bf b}\cdot{\bf u}$.
As remarked in \cite{Kraichnan:1980p248}, the invariants in  MHD do not go over smoothly into the conserved quantities of the hydrodynamics equations, except for total energy. We can expect then a different behavior, even if the magnetic field is weak initially.

The truncated MHD equations for the pair of Fourier modes $\psi_{\bf k}$ and $a_{\bf k}$, with $k\in [k_{min}, k_{max}]$, are defined in a similar way to the truncated Euler equation:
\begin{eqnarray}
\frac{\partial \psi_{\bf k}}{\partial t}&=&\frac{1}{k^2}\sum_{\bf p,q} ( {\bf p}\times{\bf q})q^2
[\psi_{\bf p} \psi_{\bf q} - a_{\bf p} a_{\bf q} ] \delta_{{\bf k},{\bf p+q}}  \label{Eq:MHDTronque2DPis}\\
\frac{\partial a_{\bf k}}{\partial t}&=&- \sum_{\bf p,q} ( {\bf p}\times{\bf q})\psi_{\bf p} a_{\bf q}\delta_{{\bf k},{\bf p+q}} \ ,\label{Eq:MHDTronque2Da}
\end{eqnarray}
with $\delta_{{\bf k},{\bf r}}$ the Kronecker delta and  with Fourier modes satisfying $\psi_{\bf k}=0,a_{\bf k}=0$ if $|{\bf k}|\ge k_{\rm max}$;
for a computational box of length $2\pi$, we have $k_{min}=1$, and with a de-aliasing using the usual 2/3 rule, $k_{max}=N/3$ where $N$ is the number of modes per dimension (we assume a box with a unit aspect ratio).

This truncated system only conserves the quadratic invariants, which can be written in Fourier space as:
\begin{eqnarray}
E&=&\frac{1}{2}\sum_{\bf k}|{\bf u}_{\bf k}|^2+|{\bf b}_{\bf k}|^2 \ , \\ 
H_{c}&=&\sum_{\bf k}{\bf u}_{\bf k}\cdot {\bf b}_{\bf -k}  \ , \\ 
A&=&\frac{1}{2}\sum_{\bf k}|a_{\bf k}|^2 \ .
\end{eqnarray}

The absolute equilibrium is the equipartition distribution of the invariants as first derived in \cite{FYFE:1976p3889} (see also \cite{Kraichnan:1980p248}), with $\alpha$, $\beta$ and $\gamma$ Lagrange multipliers, i.e. the parameters associated with the three invariants  (note that $H_C$ is not definite positive, and that $\gamma$ is a pseudo-scalar):
\begin{equation}
\alpha E+\beta A+\gamma H_c  \ .
\end{equation}
This leads to the following equilibrium spectra:
\begin{eqnarray}
E_u(k)&=&\frac{\pi k}{{\cal D}} [ k^2 \alpha +\beta  ] \label{Eq:EqAbsMHD_Ekin} \ , \\
E_b(k)&=&\frac{\pi k}{{\cal D}} k^2 \alpha \label{Eq:EqAbsMHD_Emag} \ , \\
   H_c(k)&=&-\frac{2\pi k}{{\cal D}} k^2 \gamma  \ ,\label{Eq:EqAbsMHD_Hc}
\end{eqnarray}
with
$$ {\cal D} = k^2 \left(\alpha ^2-\gamma ^2\right)+\alpha  \beta \ . $$
For small values of $\beta$,
we have equipartition of kinetic and magnetic energy $E_u(k)=E_b(k)\sim \pi k\alpha/(\alpha^2-\gamma^2)$ at all scales. 

\section{Thermalization in the absence of a uniform magnetic field} \label{s:b0=0}

 \subsection{The Orszag-Tang configuration in 2D MHD}

In order to study the thermalization of ideal MHD in two dimensions through a direct cascade of energy to small scales, we now resort to a numerical study.
The code is a standard pseudo-spectral code with periodic boundary conditions; 
the temporal scheme is a Runge-Kutta  time-stepping of fourth order that is known to accurately conserve the energy. The time steps used in our computations are given in Table \ref{Table:CompNLS-NGLR}; note that they are substantially lower than the CFL condition so that energy be preserved by the scheme for long times.

 We take as initial conditions the so-called Orszag-Tang (OT) vortex in 2D defined by the potentials \cite{ORSZAG:1979p3301}:
\begin{eqnarray}
\psi(x,y)&=&2(\cos{k_0x}+\cos{k_0y}) \ , \\
a(x,y)&=&2\cos{k_0x}+\cos{2k_0y} \ ;
\end{eqnarray}
with $k_0=1$ this vortex, made up of a neutral X-point for the magnetic field centered on the stagnation point of the velocity field, is concentrated in the large scales with $E_u=E_b=2.$ at $t=0$; hence, the initial rms values of the turbulent fields are $u_{\rm rms}\sim b_{\rm rms}\sim 2$.

The short time dynamics is identical to the one studied in \cite{FRISCH:1983p3300} at a resolution of $N^2=512^2$ Fourier modes and until the final time $T_f=1$. Here, we integrate the truncated MHD equations until $T_f=20$ using a grid of $N^2=2048^2$ points.
Note that the structures that develop at early times in physical space, namely a quadrupole in vorticity and a current dipole, were studied in detail in \cite{FRISCH:1983p3300} and are only briefly commented upon in \S \ref{s:struct}.

 \begin{figure}[h!]
\begin{center}
  \includegraphics[width=0.49\textwidth]{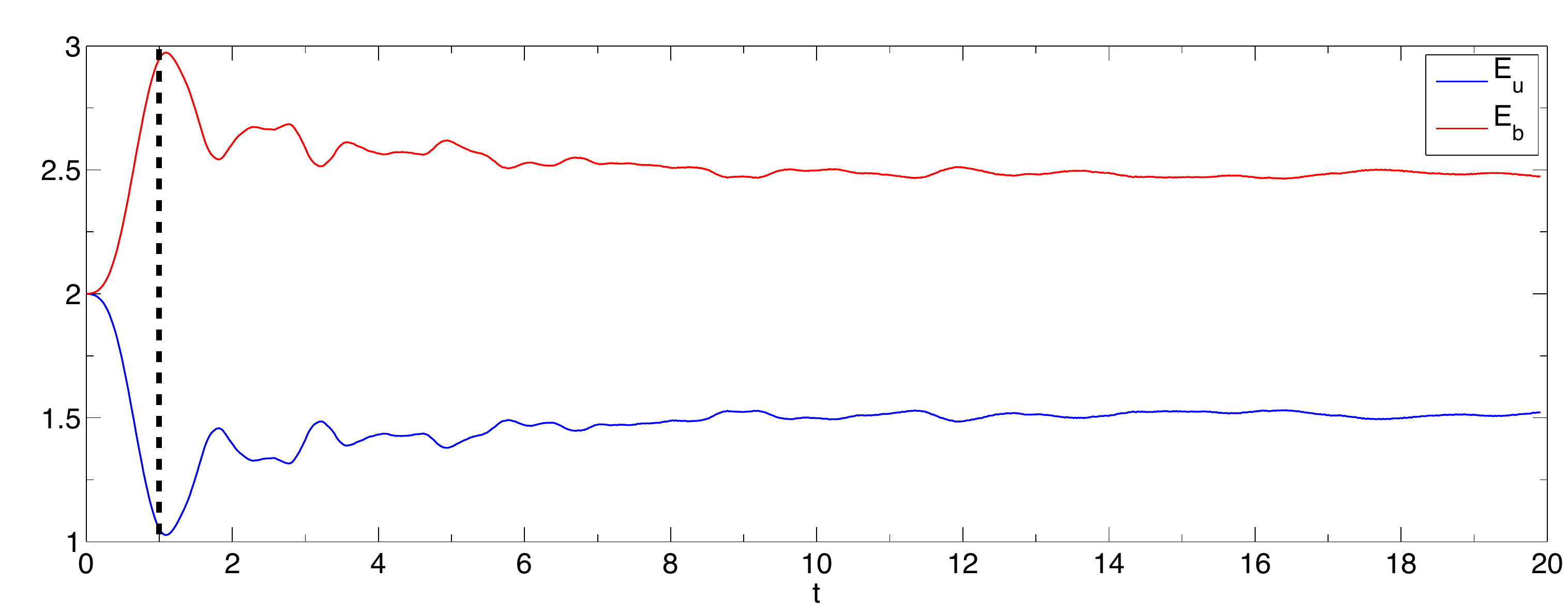}
  \includegraphics[width=0.49\textwidth]{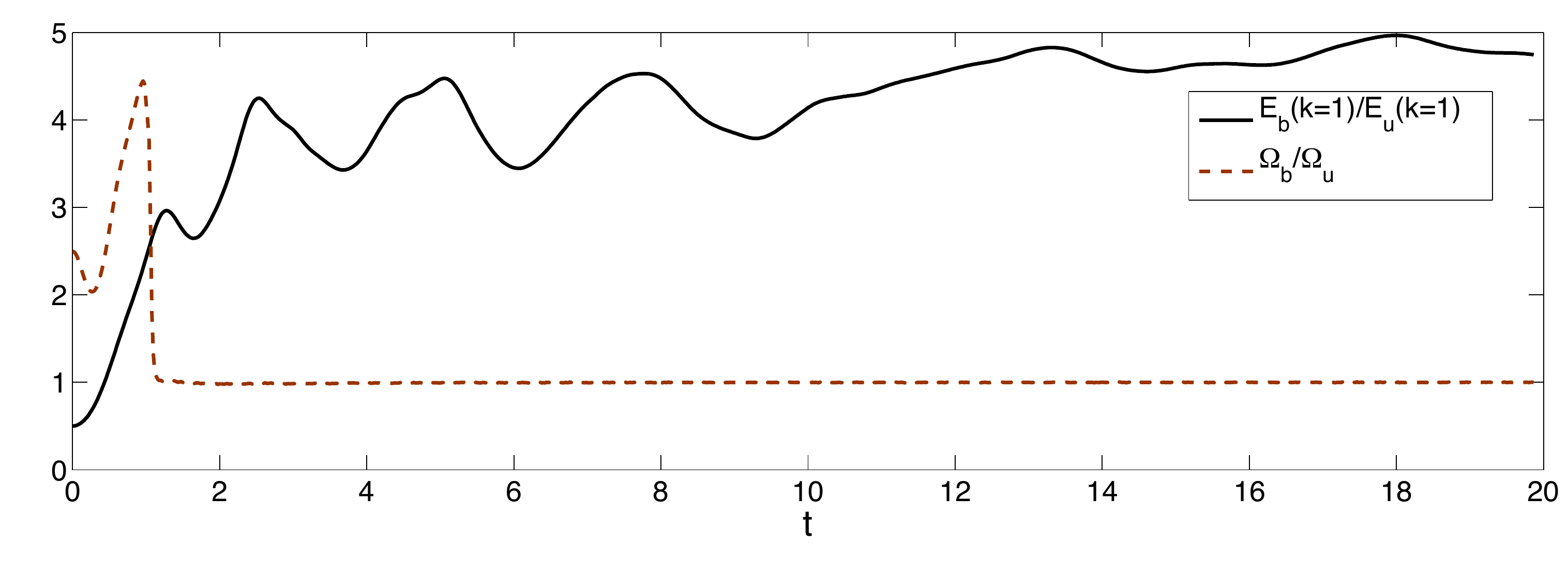}
    \caption{
    {\it Top:} Temporal evolution of  kinetic and magnetic energy, $E_u$ (in blue) and  $E_b$ (in red) for a run at a resolution of $2048^2$ grid points (Run $3$, $b_0\equiv 0$); the
        dashed vertical line indicates the final time of the numerical integration reported in  \cite{FRISCH:1983p3300} on $512^2$ grid points.
    {\it Bottom:} Temporal evolution of $E_b(k=1)/E_u(k=1)$ (solid line) and of total enstrophies $\Omega_b/\Omega_u = <j^2>/<\omega^2>$ (dash line) for the same run; as long as the large-scale modes have not equilibrated, non-linear transfer to the magnetic energy persists whereas equipartition is reached rapidly in the small scales.
    }   \label{Fig:EvoltOT} \end{center}  \end{figure}

\begin{figure*} 
\includegraphics[width=0.49\textwidth]{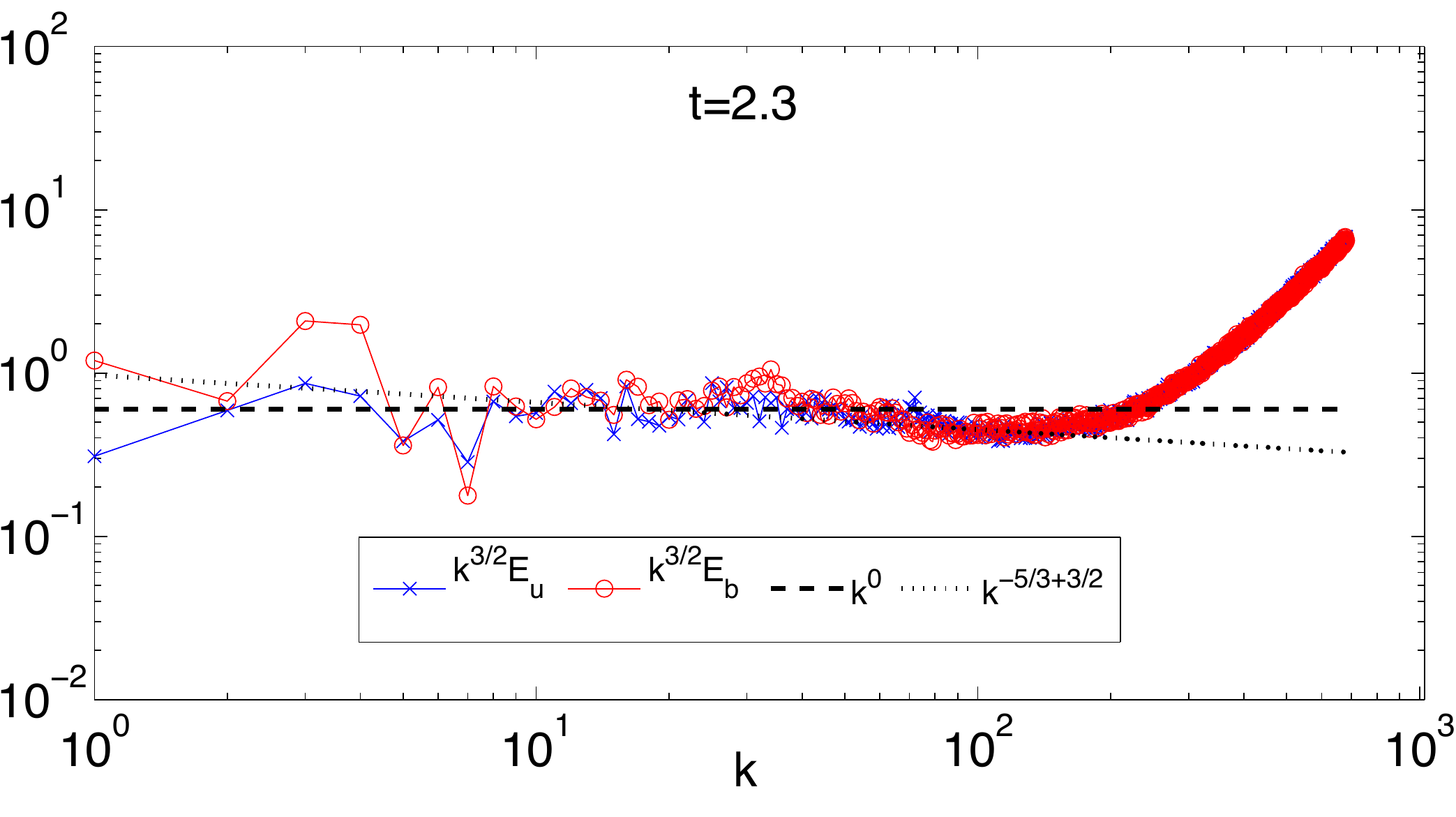}   \includegraphics[width=0.49\textwidth]{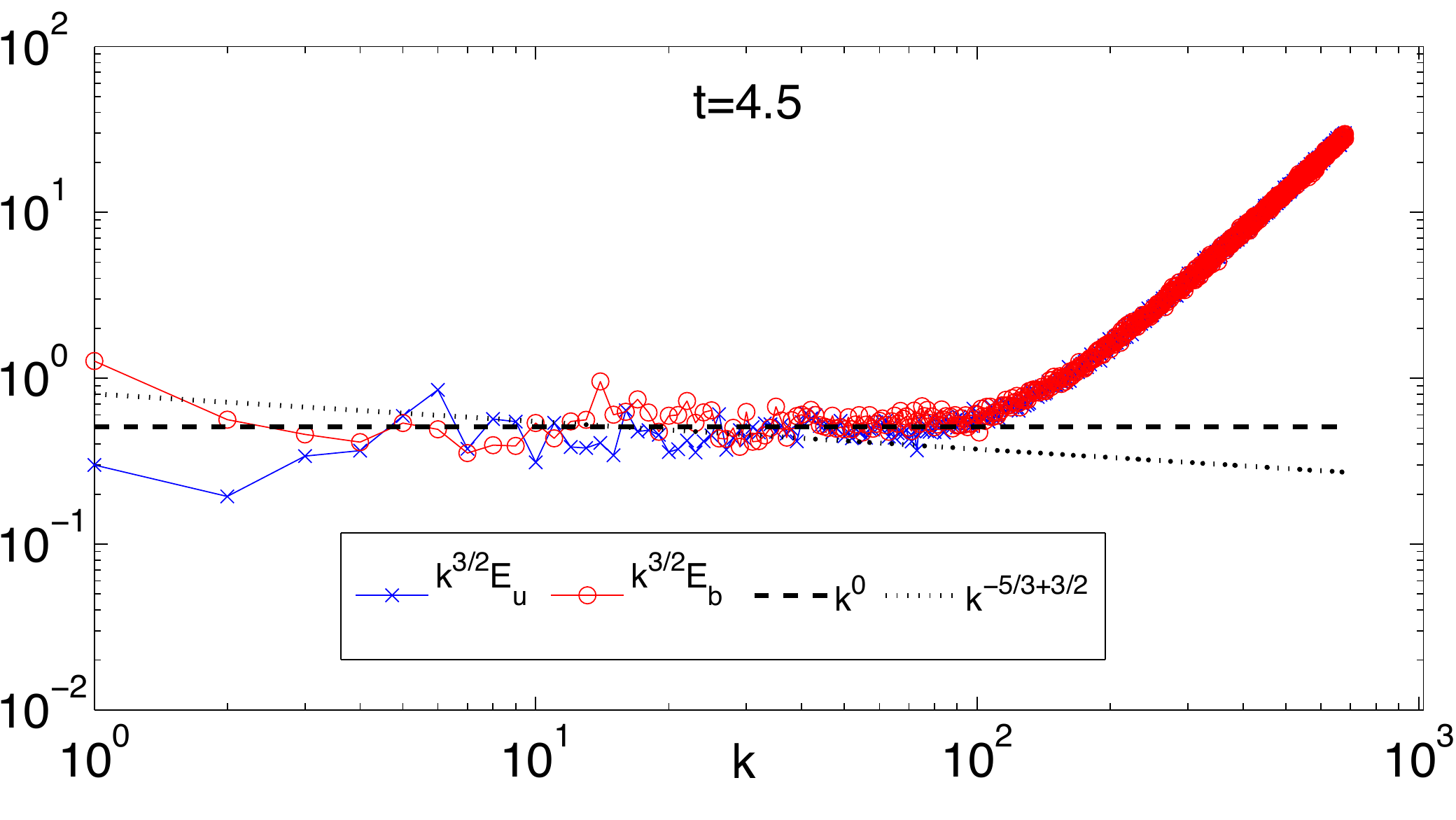}
      \includegraphics[width=0.49\textwidth]{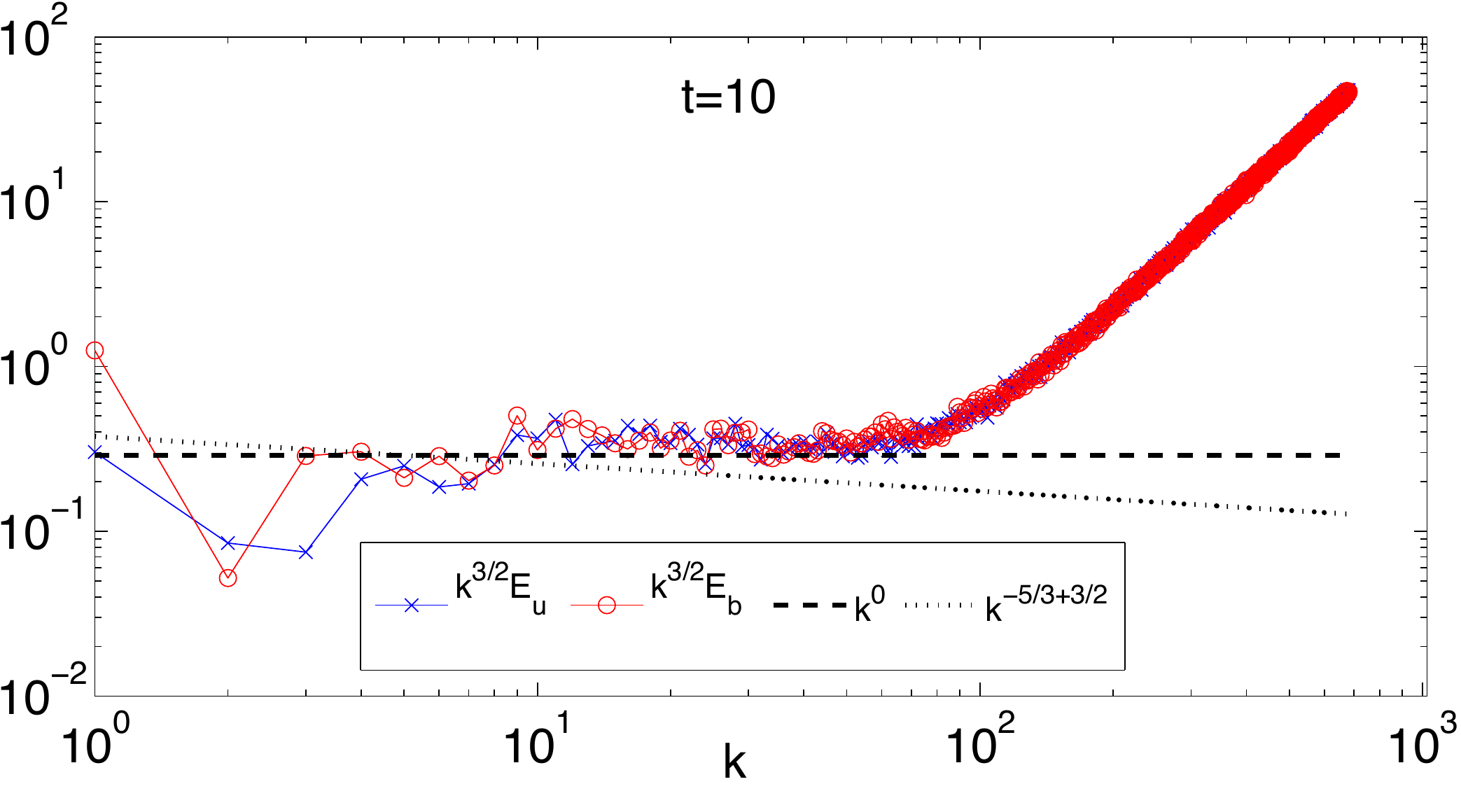}   \includegraphics[width=0.49\textwidth]{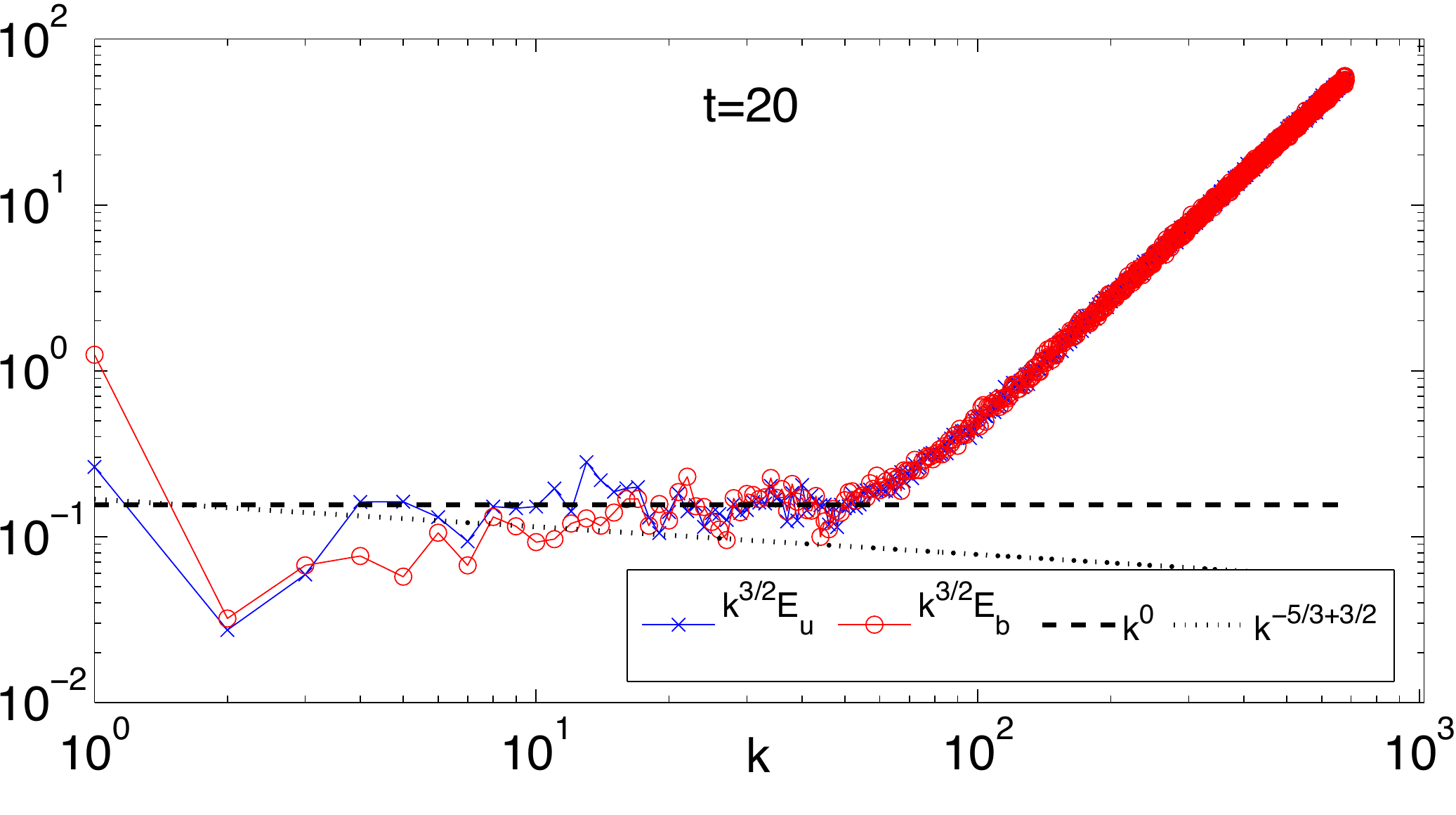}
         \caption{Energy spectra for various times, compensated by $k^{3/2}$: kinetic (x, blue) and magnetic (circle, red). Dashed and dotted lines represent fit for the K41 and IK solutions respectively, the constants appearing in these spectra being fixed at the earliest time shown, $t\approx 2.3$. Grid of $2048^2$ points, for Run $3$ with $b_0\equiv 0$ (see Table \ref{Table:CompNLS-NGLR}).
Note the fast equipartition at small scale, the IK spectrum for more than one decade at intermediate scales and intermediate times, and the persistence for long times of the domination of magnetic energy at large scale.
  }       \label{Fig:SpOT}
\end{figure*}

The temporal evolution of the kinetic and magnetic energy is shown in Fig.\ref{Fig:EvoltOT}: there is at first an energy exchange between the kinetic and magnetic energy, with the magnetic energy dominating its kinetic counterpart. As thermalization is reached, after a time $\sim 5$, the exchanges die out since the thermalized solution has no nonlinear energy flux;  $E_b/E_u$ settles at $\sim 1.7$, leading to $u_{\rm rms}\sim 1.5$, $b_{\rm rms}\sim 2.5$. Such a moderate  excess of magnetic energy is often observed in the Solar Wind \cite{bruno}; it corresponds to a slight departure from equipartition that can be attributed, in the context of ideal flows, as due to the effect of the conservation of the magnetic potential, see Eqs.(\ref{Eq:EqAbsMHD_Ekin}, \ref{Eq:EqAbsMHD_Emag}). 
We define as usual the total energy and total enstrophy spectra by summing the basic fields
on circular shells of width $\Delta k = 1$:
\begin{eqnarray}
E_T(k,t)&=& {\frac1 2} \sum_{k_-< |{\bf k'}| <  k _+} [ |{\bf \hat u}({\bf k'},t)|^2 + |{\bf \hat b}({\bf k'},t)|^2 ]   \, \\                             
\Omega_T(k,t)&=& {\frac1 2} \sum_{k_- < |{\bf k'}| <  k_+ } [ |{\bf \hat \omega}({\bf k'},t)|^2 + |{\bf \hat j}({\bf k'},t)|^2  \, ,\label{eq_energy}
 \end{eqnarray}
with $k_{\pm}= k \pm \Delta k/2$\ ; ${\bf \hat{u}}({\bf k,t})$, 
${\bf \hat \omega}({\bf k},t)$, ${\bf \hat{b}}({\bf k,t})$ and ${ \hat{j}}({\bf k,t})$ are the Fourier transforms of the velocity, vorticity, magnetic field and of the current density, with $j_z=\nabla \times {\bf b} \hat e_z$. In Fig. \ref{Fig:EvoltOT} (bottom), the ratio of the kinetic and magnetic energy spectra at $k=1$ is displayed, as well as the ratio of $<j^2>/<\omega^2>$ (corresponding to the ratio of kinetic and magnetic dissipation when equal viscosity and resistivity are re-introduced in the equations).
Observe that magnetic energy dominates at large scale, a sign of a plausible inverse cascade associated with the magnetic potential and confirming that the defect in equipartition is indeed due to the large-scale behavior of the system, since almost exact equipartition is observed at small scale (see e.g. Fig.\ref{Fig:SpOT}).

\begin{table*} 
 \begin{tabular}{|c|c|c|c|c|c|c|c|} \hline
Runs & $N^2$, $T_f$ & ${\bf b_0=0}$ & ${\bf b_0=2}$& ${\bf b_0=4}$ & ${\bf b_0=8}$&${\bf b_0=12}$ & ${\bf b_0=16}$\\
\hline \hline
1 & \ $256^2$, $T_f=500$  & $-$   & $-$ & $-$& $-$  & $-$ & $dt=1/16000$ \\
2 & \ $512^2$, $T_f=60$  & $dt=1/1600$   & $dt=1/3200$ & $dt=1/3200$& $dt=1/6400$  & $dt=1/8000$ & $dt=1/8000$ \\
3 & $2048^2$, $T_f=20$  & $dt=1/20000$   & $-$ & $-$& $-$  & $-$ &$-$ \\
4 & $2048^2$, $T_f=5$   & $-$   & $dt=1/10000$ & $dt=1/20000$& $dt=1/40000$  & $-$ & $dt=1/40000$ \\
5 & $4096^2$, $T_f=1.2$   & $dt=1/20000$ & $-$& $-$  & $-$ & $-$& $-$ \\
\hline \end{tabular}
\caption{Time step $dt$ for the various ideal MHD numerical simulations presented in this paper. In all cases, the total energy is conserved at better than 0.1\% at the final time of the computation $T_{\rm f}$. Note that $b_{\rm rms}\sim 2.5$ when $b_0\equiv 0$, so that all cases have a dominant imposed field $b_0$, except the one with $b_0=2$ which can be viewed as an intermediate case.
}
 \label{Table:CompNLS-NGLR} \end{table*}

\begin{figure} 
\includegraphics[width=8.5cm]{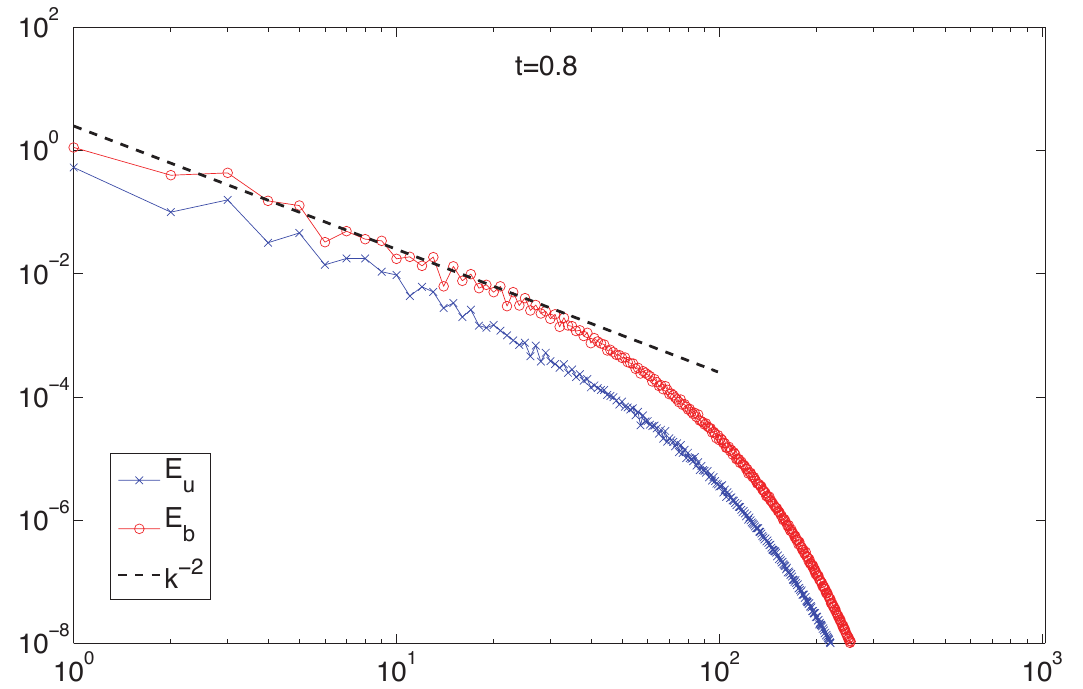} 
         \caption{Kinetic (x, blue) and magnetic (circle, red) energy spectra (color online) at $t\sim 0.83$, in the early phase of the flow for Run $3$ with $b_0\equiv 0$. The black dashed line corresponds  to $k^{-2}$, a spectrum expected in the presence of quasi-singular structures.  Note that magnetic energy dominates at all scales at that time.
         }\label{f:early} \end{figure}

Concerning spectral variations, at early times as displayed in Fig.\ref{f:early}, the spectra show a clear $k^{-2}$ dependence, that can be associated with the formation of quasi-singular current and vorticity sheets \cite{FRISCH:1983p3300}. This behavior is exhibited by all spectra (kinetic, magnetic and total energy). 
It is also present (data not shown) when a uniform magnetic field is imposed to the flow, granted its magnitude is not too large compared to the rms values of the fluctuating field: for $b_0=0.5$, a similar spectrum obtains whereas for $b_0=4$, the spectrum is much steeper, due in part to a slower dynamical evolution (see next Section). 
This $k^{-2}$ spectrum associated with sharp current and vorticity sheets persists as long as convergence of the partial differential equations is assured, i.e. until the thickness of current and vorticity structures is comparable (but still larger)
than the grid size (see also the discussion concerning the logarithmic decrement displayed in Fig. \ref{Fig:delta}). 

It was shown in \cite{Cichowlas:2005p1852} in the context of three-dimensional (D=3) Euler flows that, as time evolves, another dynamics takes place: once the flow behaves as a truncated system of modes and thermalization begins, the modes following a $k^{D-1}$ law at small scale for the three-dimensional Euler equations produce an eddy viscosity for larger scales; a quasi-turbulent regime follows at intermediate scales and intermediate times (i.e. before thermalization occurs everywhere), with an inertial range close to a Kolmogorov law corresponding to the forced dissipative case.

Similarly in the present $2{\rm D}$ case, once the smallest resolved scale has been reached, the overall solution becomes noisier, the vorticity and current sheets curve and interact, and the Fourier spectra evolve as well (structures are displayed in Fig.\ref{Fig:Struc} below).

The kinetic and magnetic energy spectra, compensated by a $k^{3/2}$ power law, are shown in Fig.\ref{Fig:SpOT} for several times, in the range $t\sim 2.5$ to $t\sim 20$.
As in the case of thermalization for the truncated Euler equation evolution in hydrodynamics with large-scale initial conditions, a clear scale separation now appears in these spectra. 

It is possible then to define a wavenumber $k_{\rm th}(t)$ where the thermalized $k^1$ power-law zone starts (numerically determined by seeking the minimum of $E(k)$). The total thermalized energy and enstrophy are thus defined as:
\begin{equation}
{E}_{\rm  th}(t)=\sum_{k_{\rm th}(t)}^{ k_{\rm max}}E_T(k,t)\,,\hspace{0.5cm}{\Omega}_{\rm  th}(t)=\sum_{k_{\rm th}(t)}^{ k_{\rm max}}\Omega_T(k,t).\label{Eq:Eth}
\end{equation}
We now can estimate the effective energy dissipation rate as $\varepsilon_{\rm th}(t)=\frac{d E_{\rm th}}{dt}$. Similar to the $3D$ Euler case \cite{Cichowlas:2005p1852,Krstulovic:2009p4422}, this quantity presents a maximum, here for $t\approx 2.5$ (data not shown). Note that this time also corresponds to the interval of time during which the total dissipation, when viscosity and magnetic resistivity are added to the primitive equations, is quasi-stationary (see \cite{POLITANO:1989p3134}). The scaling laws for a K41 spectrum (dotted line), and for the IK spectrum (dashed line) defined respectively as:
\begin{equation}
E_u^{K41}(k)\approx E_b^{K41}(k)\sim C_{K}\varepsilon^{2/3}k^{-5/3} \ , \label{Eq:K41LawMHD}
\end{equation}
and
\begin{equation}
E_u^{IK}(k)\approx E_b^{IK}(k) =C_{IK}[\varepsilon b_{\rm rms}]^{1/2}k^{-3/2} \label{Eq:IKLawMD}
\end{equation}
 are also displayed in Fig. \ref{Fig:SpOT} at large scales, with $b_{\rm rms}\sim 2.5$ the root mean square value of the fluctuating magnetic field.
  The numerical constants $C_K$ and $C_{IK}$ appearing in front of the K41 and IK spectra are evaluated (see Fig. \ref{Fig:SpOT}.a)
  by fitting the  large-scale part of the spectra at an early time after the first Alfv\'enic energetic exchange, $t\sim 2.5$  (note that the turbulence spectrum is already visible at $t\sim 2.0$, not shown). We find for these constants $C_{K}=2$ and $C_{IK}=0.8$ at the time of the maximum of the effective dissipation $\varepsilon_{\rm th}(t)$. 
  
Therefore, we can safely conclude, on the basis of the examination of the results plotted in Fig. \ref{Fig:SpOT} that the IK spectrum is obtained in the intermediate spatio-temporal range of ideal dynamics in MHD in two dimensions for the OT vortex. Also note the good equipartition of kinetic and magnetic energy in the thermalized range, and the domination of magnetic energy at large scale, again a sign of a plausible inverse cascade associated with the magnetic potential.

 \subsection{Random initial conditions}
 
\begin{figure*} 
\includegraphics[width=0.49\textwidth]{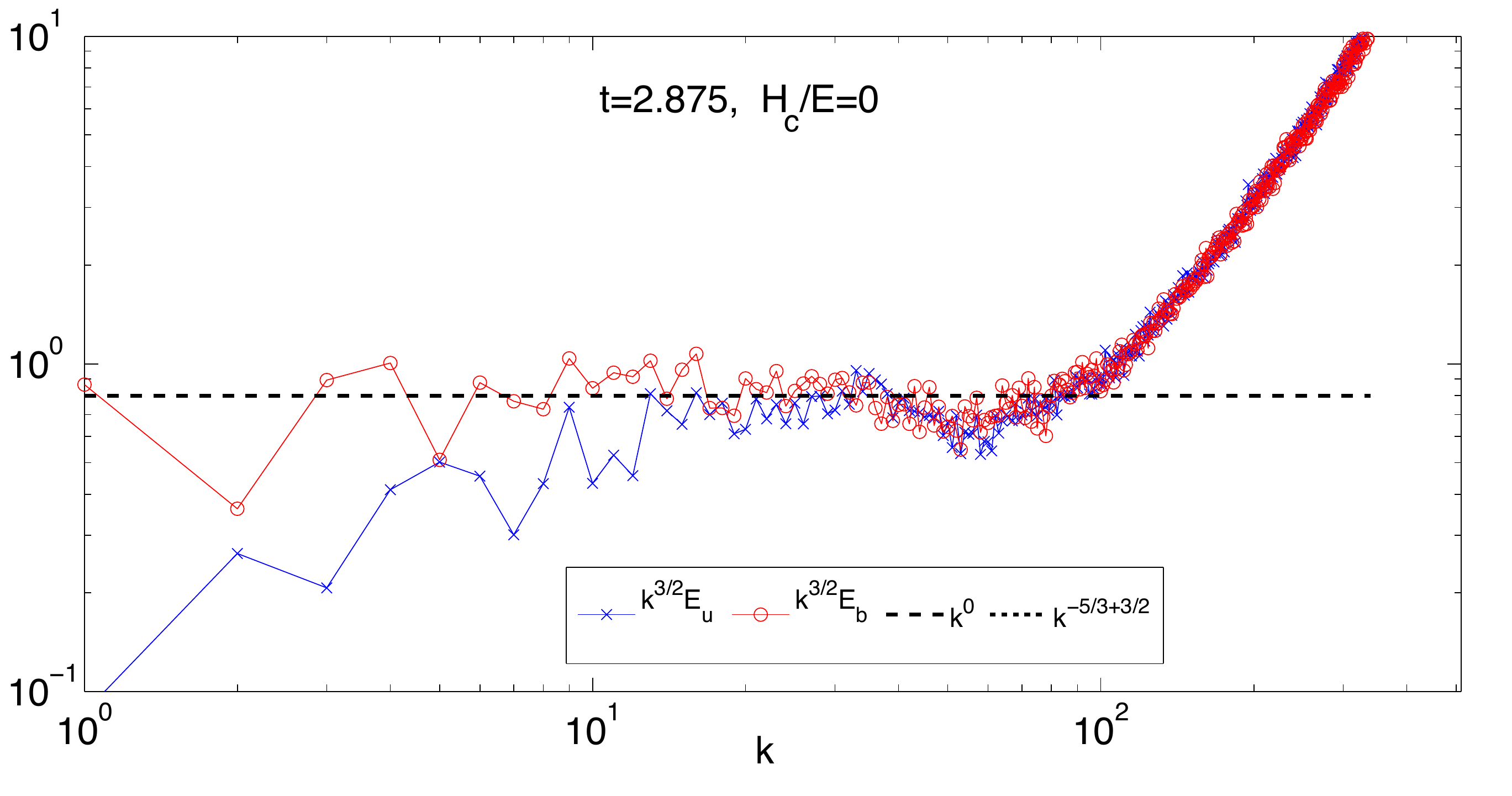}   \includegraphics[width=0.49\textwidth]{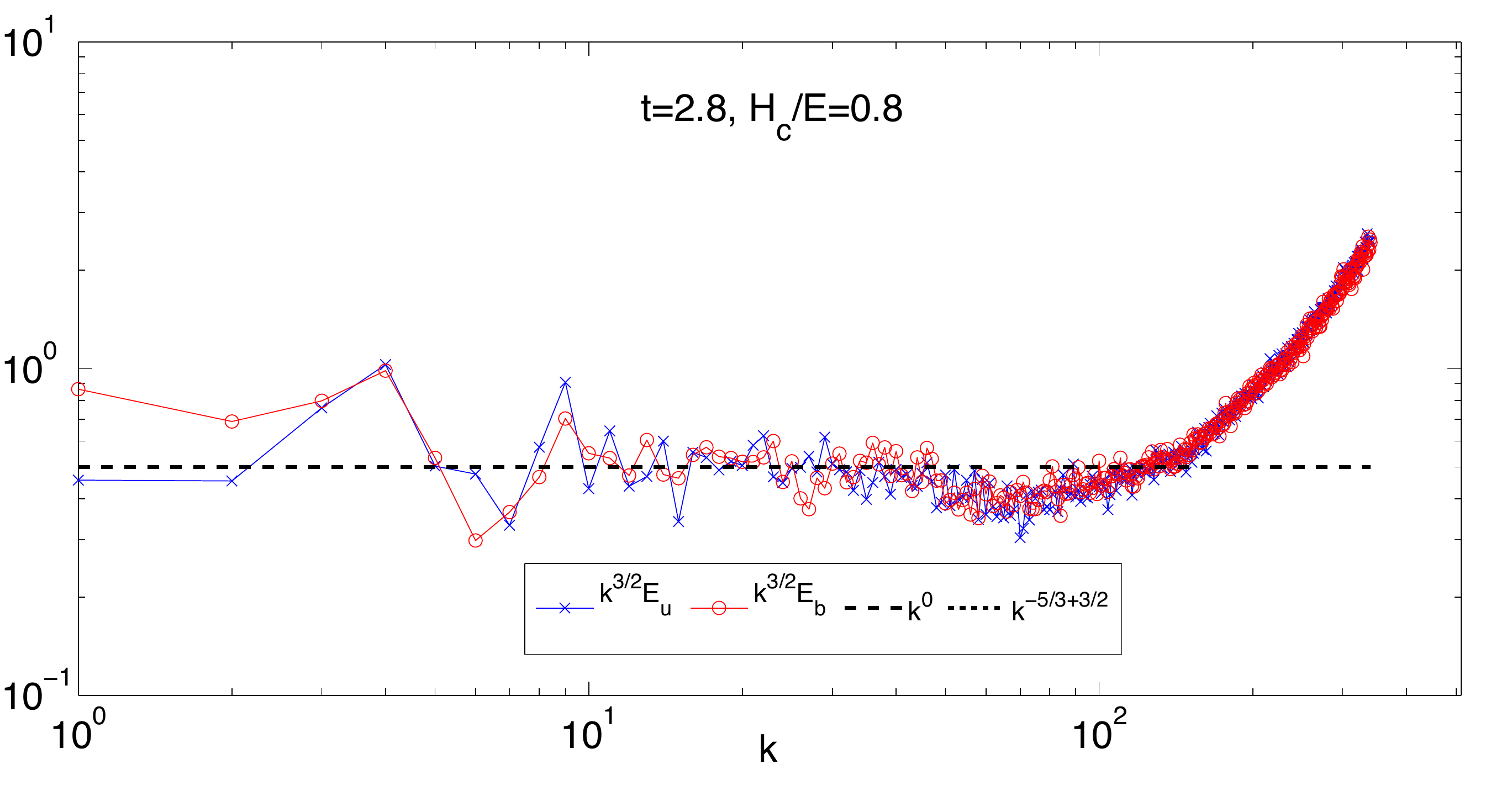}
 \caption{Energy spectra for random initial conditions with normalized velocity-magnetic field correlation of $0$ (left) and $0.8$ (right), $3/2$-compensated kinetic (x, blue) and magnetic (circle, red). The dotted lines correspond to an IK spectrum, with $C_{IK}=0.8$ (left) and $C_{IK}=0.5$ (right), see eq. (\ref{Eq:IKLawMD}).
}  \label{Fig:RandCondIni}  \end{figure*}

Similar results obtain for random initial conditions, as can be seen in Fig.\ref{Fig:RandCondIni} plotting spectra in that case for a normalized global velocity-magnetic field correlation ($H_c/E$) that takes values of 0 (left) and 0.8 (right): a thermalized spectrum is observed at small scale, and a turbulent IK-spectrum at large-scale. 
However, note that non-universality has been obtained in MHD \cite{lee2} in the decaying dissipative case;  a K41 spectrum was also previously obtained in decaying 2D MHD turbulence with random initial conditions using a Lagrangian model \cite{mininni_LAMHD2D}.

 \subsection{The pseudo-dissipative range, or lack thereof}

Remark that the end of the pseudo-inertial range is not followed by a sharp decrease in energy before the thermalized $E_T(k)\sim k$ small-scale spectrum, contrary to the 3D Euler case \cite{Cichowlas:2005p1852,Krstulovic:2009p4422}. However, computing a dissipation wavenumber using the IK spectrum and the corresponding evaluation of the transfer time that leads to the IK spectrum, gives 
 $\ell_{diss}\sim \nu^{2/3}$, where in our ideal case the viscosity is replaced by a turbulent expression based on the thermalized energy, 
$\nu_{turb}\sim \sqrt{E_{th}}/k_{max}\sim 1/k_{max}$. Similarly, one can recall that the eddy viscosity computed with the EDQNM closure gives a non-zero contribution \cite{Pouquet:1978p2787}. However, several remarks are in order.
First of all, the small-scale velocity leads to no contribution to an eddy viscosity for the dynamics of the large-scale velocity field \cite{Kraichnan:1980p248}, so the sole contribution to dissipation of $E_u$ will stem from the small-scale magnetic field. Furthermore, the eddy resistivity contributions of the small-scale velocity and magnetic field exactly compensate each other when $E_u=E_b$ in the small scales, which is the case here (with small $\beta$). However, one can compute the correction to equipartition which is known to follow a $k^{-2}$ law in the dissipative case \cite{PFL}; therefore, one could expect a non-zero contribution to turbulent  viscosity in 2D MHD as well. Finally, another argument can be put forward to explain the lack of sharp decrease of the spectra before the thermalized range, namely that there are not enough modes in quasi-equilibrium to produce a sufficient amount of effective dissipation. Indeed, in the 3D case, it was shown that a number of roughly $256^3$ modes was necessary to see this internal decrease of the energy spectrum before the thermalized spectrum; this would correspond to a computation on a grid of $4092^2$ points in 2D. However, observe that using a resolution of $2048^2$ points, a slight dissipative zone seems to appear at $t=2.3$
(for $k\approx 50$).
 This point will await further study.
 
To try to understand further the lack of dissipation range, we also resorted to an examination of the two-dimensional Euler case that is given in some detail in Appendix \ref{Sec:Euler}. The truncated Euler equations relax toward the statistical equilibrium in an analogous way to the three dimensional case. The main difference is the presence of a direct cascade of enstrophy. This quantity plays the role of the energy in $3D$, thermalizing in equipartition at large wave-number and yielding a $k^{-1}$ law in the inertial zone. Again, a remarkable difference with the $3D$ Euler case is the absence of a dissipative zone that is due here to a vanishing $2D$ eddy viscosity.

\section{Dynamical slowing down in the presence of a uniform magnetic field} \label{s:b0not=0}
 
 In the presence of a strong imposed uniform magnetic field of amplitude $b_0$, it is known that the dynamics is slowed down, including in the ideal case  \cite{FRISCH:1983p3300}; this is in fact at the basis of the argument of Iroshnikov and Kraichnan for an energy spectrum in MHD different from the Kolmogorov spectrum, and it is the feature on which the weak turbulence development for MHD relies upon \cite{Galtier:2000p3795}, using the smallness of the ratio of the Alfv\'en time to the eddy turnover time 
 \begin{equation}
 R_{WT}=\tau_A/\tau_{NL} \ ,
 \label{RT} \end{equation}
  with $\tau_A=L_0/b_0$ and $\tau_{NL}=L_0/U_0$, $L_0$ and $U_0$ being respectively the characteristic large scale and velocity of the flow. It is also claimed in  \cite{FRISCH:1983p3300}, in the framework of ideal 2D MHD, that in fact the development of small scales in the presence of a sufficiently strong ${\bf b}_0$ is arrested, with a smallest excited scale $\ell_B$, that depends on ${\bf b}_0$ and that can be larger than the smallest resolved scale of the flow in a computation at a given resolution.
 
  In view of the increased power of computers available today, we revisit the effect a uniform field has on the formation of small scales in the ideal case of $2D$ MHD. Compared to the work in \cite{FRISCH:1983p3300}, we are now performing computations for longer times, for different values of the magnetic field $b_0$ and for higher resolutions, using here grids up to $N^2=4096^2$ points. 
  Note that, for a magnetic field aligned with the $x$-axis,  eqs.(\ref{Eq:MHDPsipot}-\ref{Eq:MHDapot}) are modified by performing the substitution $a\to a+b_0y$. Remark that now the square magnetic potential $A=<a^2>$ is no longer conserved and the absolute equilibrium spectra given in Eqs.(\ref{Eq:EqAbsMHD_Ekin}-\ref{Eq:EqAbsMHD_Hc}) are obtained in this case by setting $\beta=0$.
 
\begin{figure} 
\includegraphics[width=0.49\textwidth]{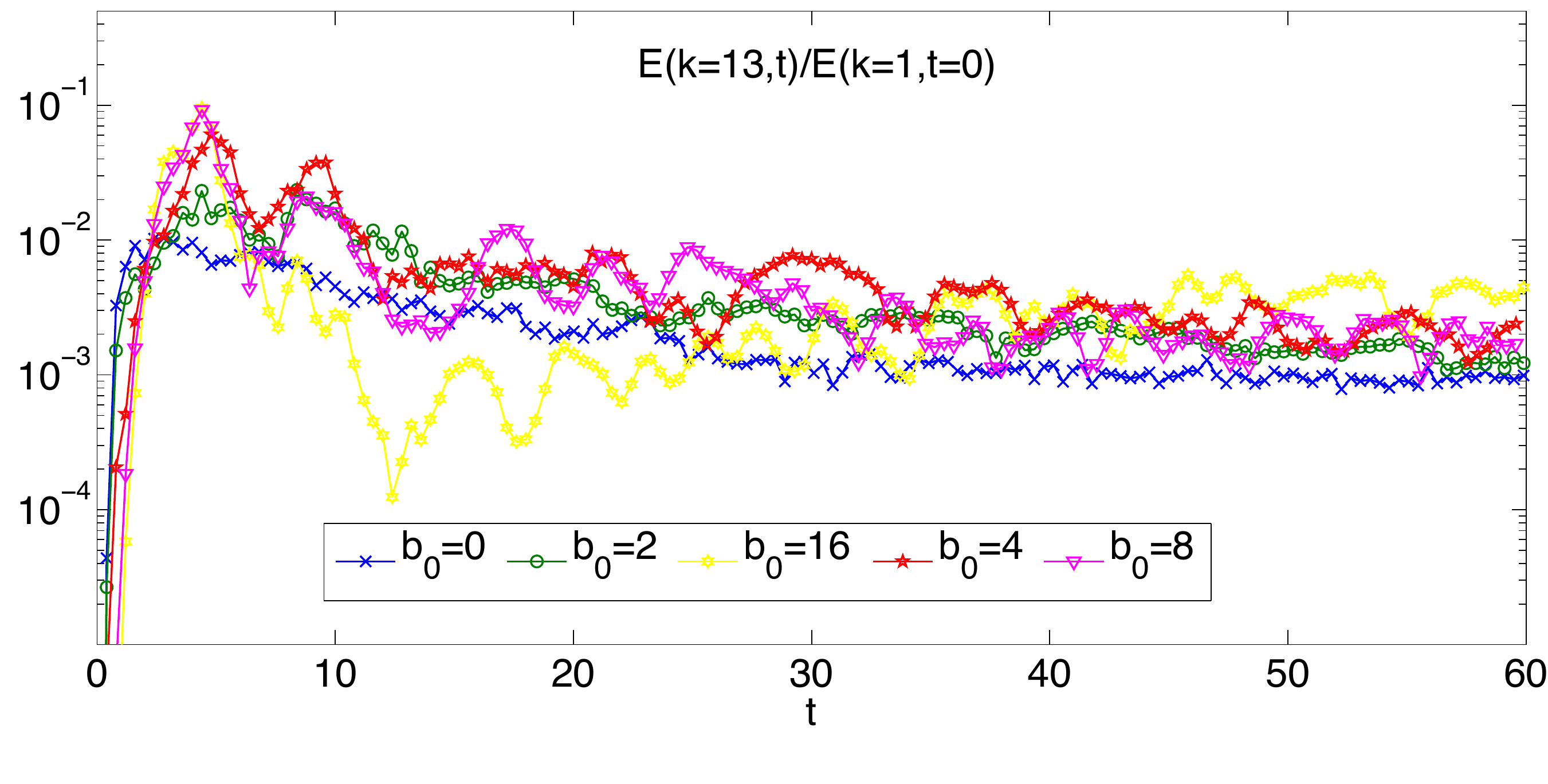} 
\includegraphics[width=0.49\textwidth]{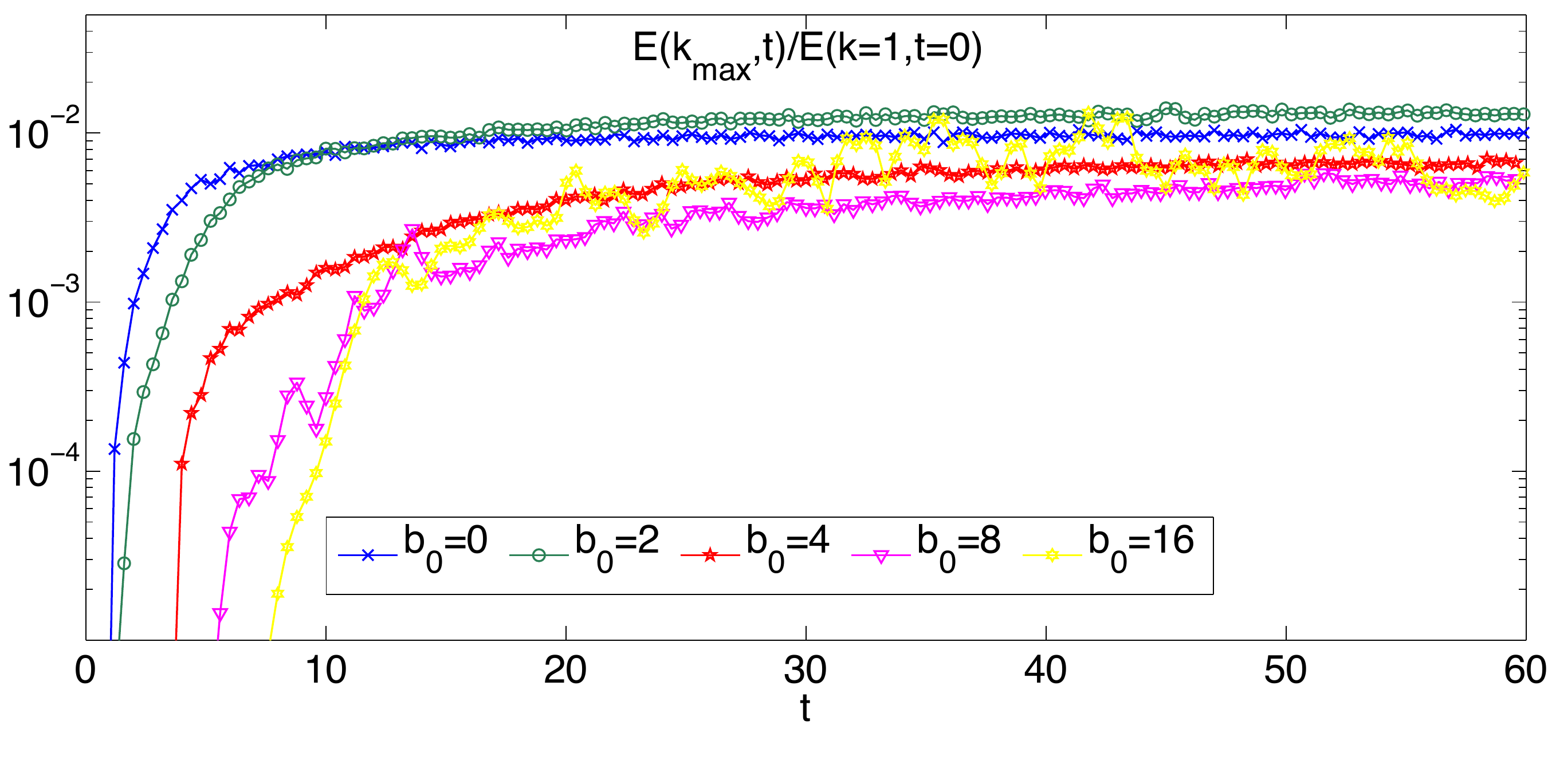} 
         \caption{Temporal evolution of total (kinetic plus magnetic) energy in mode $k=13$ (top) and $k=k_{max}$ (bottom) for several imposed mean fields: 
  $b_0=0$ (x), $b_0=2$ (circle), $b_0=4$ ($\ast$), $b_0=8$ (triangle) and $b_0=16$ (star); both modes are normalized by the initial value of the energy in the gravest mode.
         Note the substantial delays in the evolution as $b_0$ increases, and the final saturation level. Grids of $512^2$ points (Runs 2).
}\label{f:modes} \end{figure}

 In Fig.\ref{f:modes} we give the temporal evolution of two modes, one in the middle of the resolved range ($k=13$), and the other one at the end ($k=k_{max}=N/3$), both normalized by E(k=1,t=0), and for several values of the imposed field (see caption). We observe a delay in the early dynamics of the modes as $b_0$ increases, followed at long times by a saturation once equilibrium is reached (particularly so for $E(k_{max})$).
  
\begin{figure} 
\includegraphics[width=0.45\textwidth]{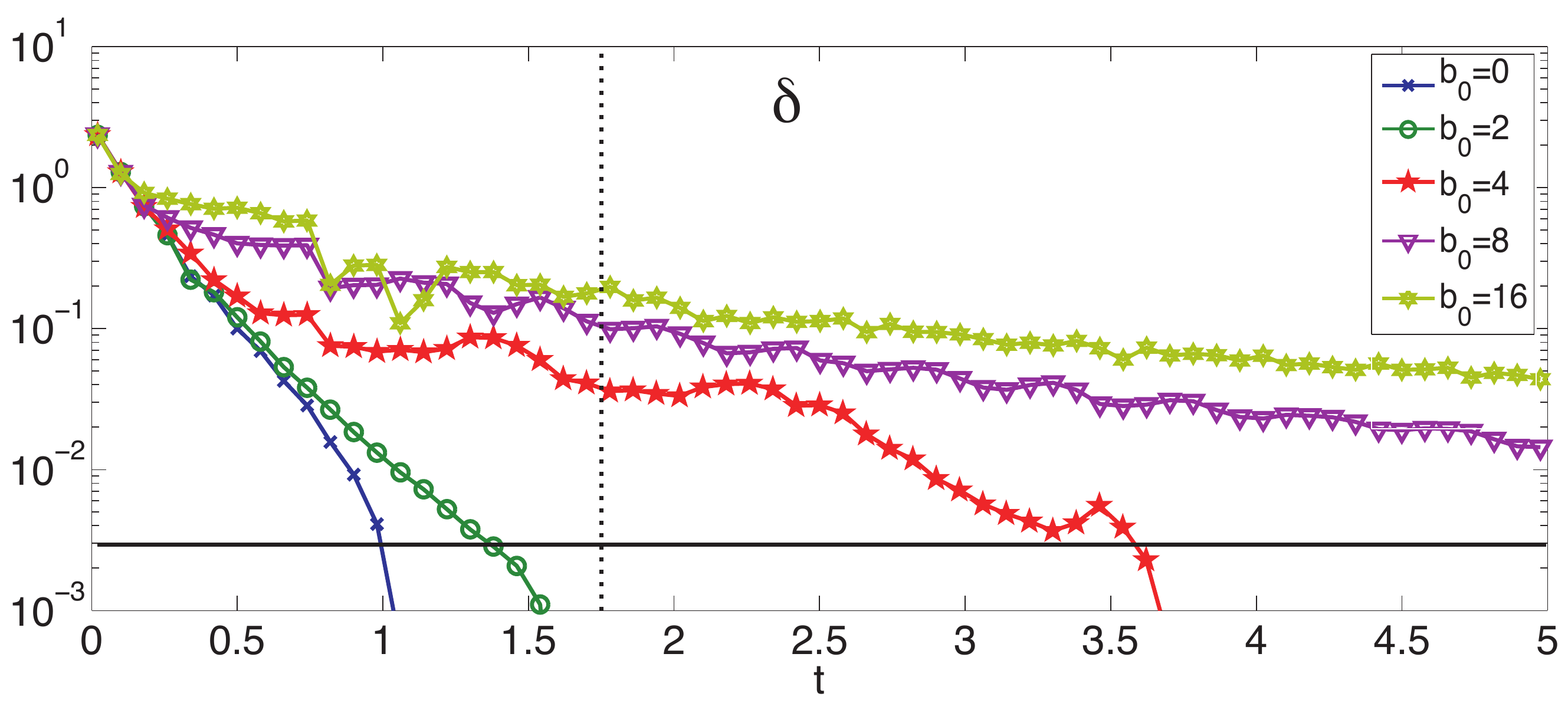} 
\includegraphics[width=0.45\textwidth]{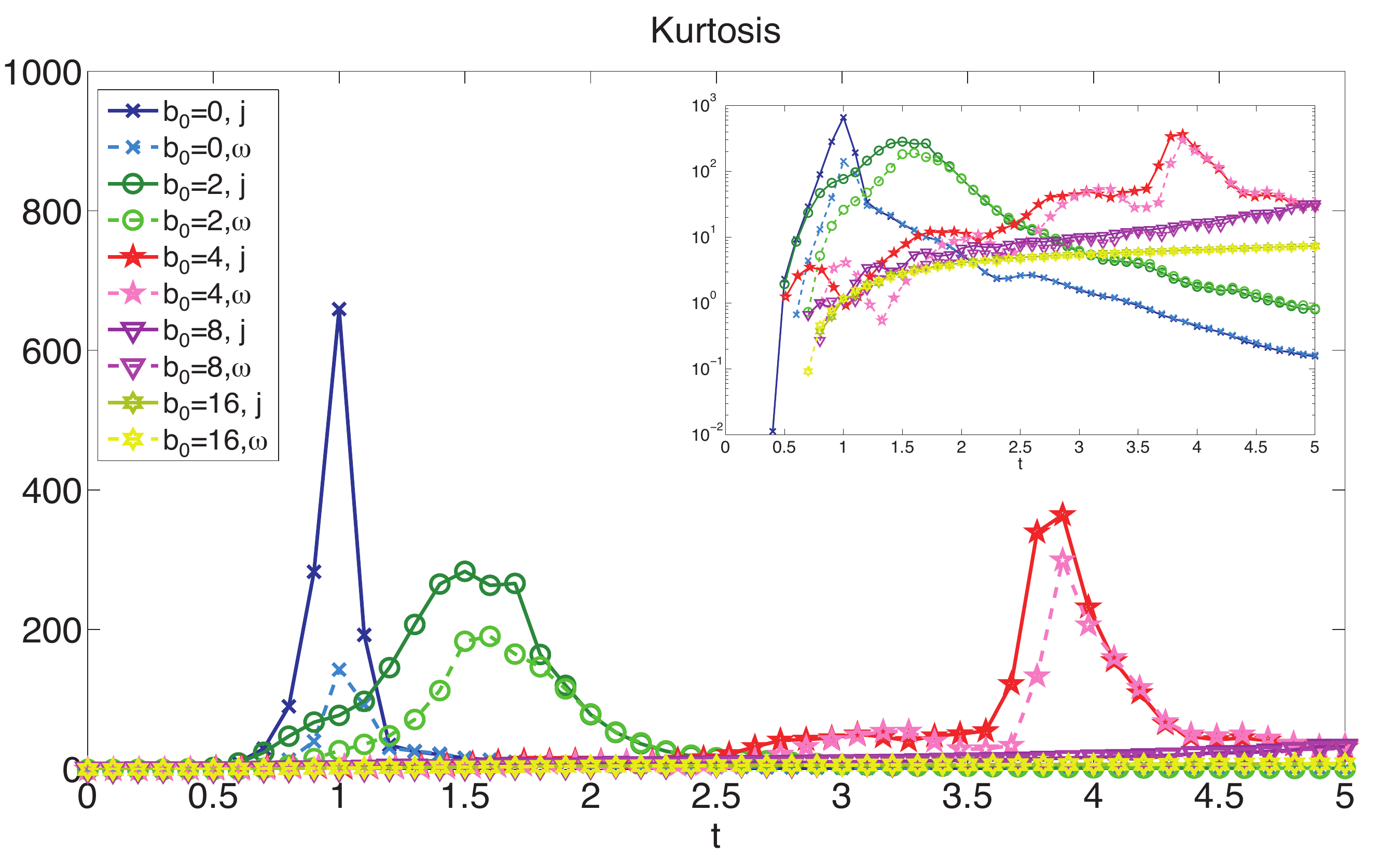}
         \caption{
         {\it Top:} Temporal evolution of the logarithmic decrement $\delta(t)$ with $b_0  =0$ (x), $b_0  =2$ (circle), $b_0=4$ (star), $b_0=8$ (triangle) and $b_{0}=16$ ($\ast$). Resolution of $2048^2$ grid points, Runs 3 and 4. The horizontal solid line gives the smallest admissible value of $\delta$ for the PDEs to be well resolved for these grids, and the vertical dashed line represents the time at which integration was stopped in  \cite{FRISCH:1983p3300} for a computation on a grid of $256^2$ points. No saturation of the temporal evolution of the logarithmic decrement $\delta(t)$ is observed provided the numerical resolution allows one to compute for long enough times. 
         {\it Bottom:} Kurtosis $\gamma_2=\kappa_4/\kappa_2^2$ (with $\kappa_i$ the cumulant of order $i$) for the same runs as above, of the current $j$ and vorticity $\omega$ for different values of the background constant magnetic field; the current is shown with a solid line (and darker color), whereas the vorticity is given with a dash line (and lighter color, see left inset). In the right inset is shown the same plot in $log-lin$ coordinates; 
 }\label{Fig:delta} \end{figure}

The logarithmic decrement technique \cite{Sulem:1983} is now applied to quantify further the delay of the onset of the evolution as the amplitude of the imposed uniform field is increased. If the fields are regular, then the energy spectra must decay at least exponentially at large wave-number $k$.  Based on this assumption, the logarithmic decrement  $\delta(t)$ is defined by the large $k$ asymptotic of the energy spectra:
\begin{eqnarray}
E_T(k)=c(t)k^{-m(t)} e^{-2 \delta(t)k}\label{Eq:delta0} \ ;
\end{eqnarray}
$\delta(t)$ is measured by fitting the long wave-number range and the minimum admissible value is determined by the relation $\delta(t)k_{\rm max}=2$ ($\delta=$twice the mesh).
The temporal evolution of $\delta(t)$ for different values of the imposed magnetic field is displayed in Fig.\ref{Fig:delta} in a \emph{log-lin} plot. 
 
The presence of a strong magnetic field slows down the nonlinear interactions and $\delta(t)$ remains above the minimum admissible value for a longer time the stronger the value of ${\bf b}_0$, but it finally reaches the condition $\delta(t)k_{\rm max}=2$ near $t=3$ for $b_0=4$ (and $t\sim 5.8$ for ${\bf b}_0=8$, see Fig.\ref{Fig:delta}). From this study, we can deduce that the system can reach statistical equilibrium even in the presence of a strong imposed magnetic field, but that the convergence toward such a state is considerably hampered.
In Fig.\ref{Fig:delta} (bottom), we also display the temporal evolution of the kurtosis for the current and vorticity and for several imposed uniform magnetic fields. Again, the delay in the formation of small scales is observed, as well as a tendency towards Gaussianity once the thermalization takes place.

 \begin{figure*} 
\includegraphics[width=0.49\textwidth]{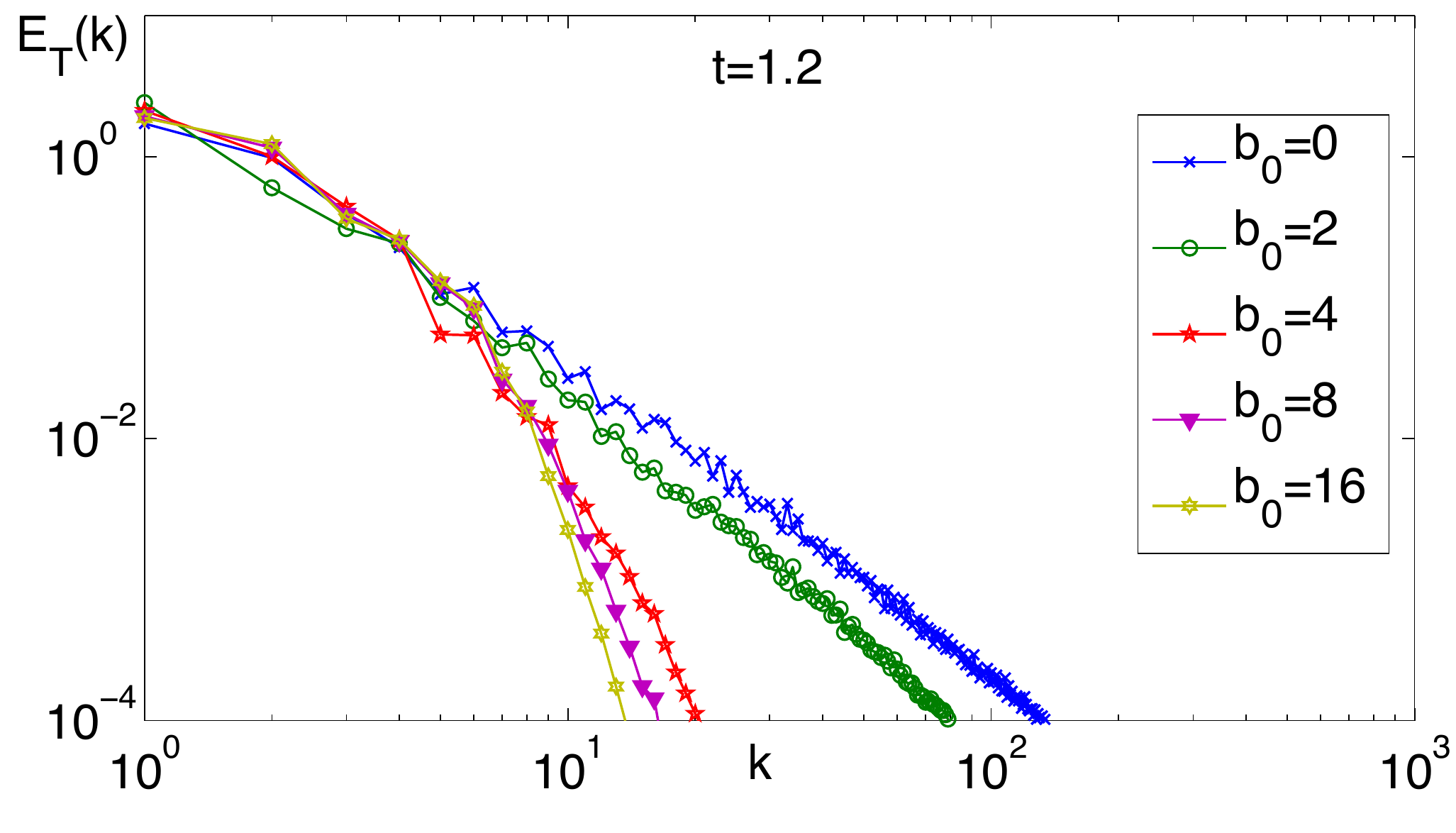}  \includegraphics[width=0.49\textwidth]{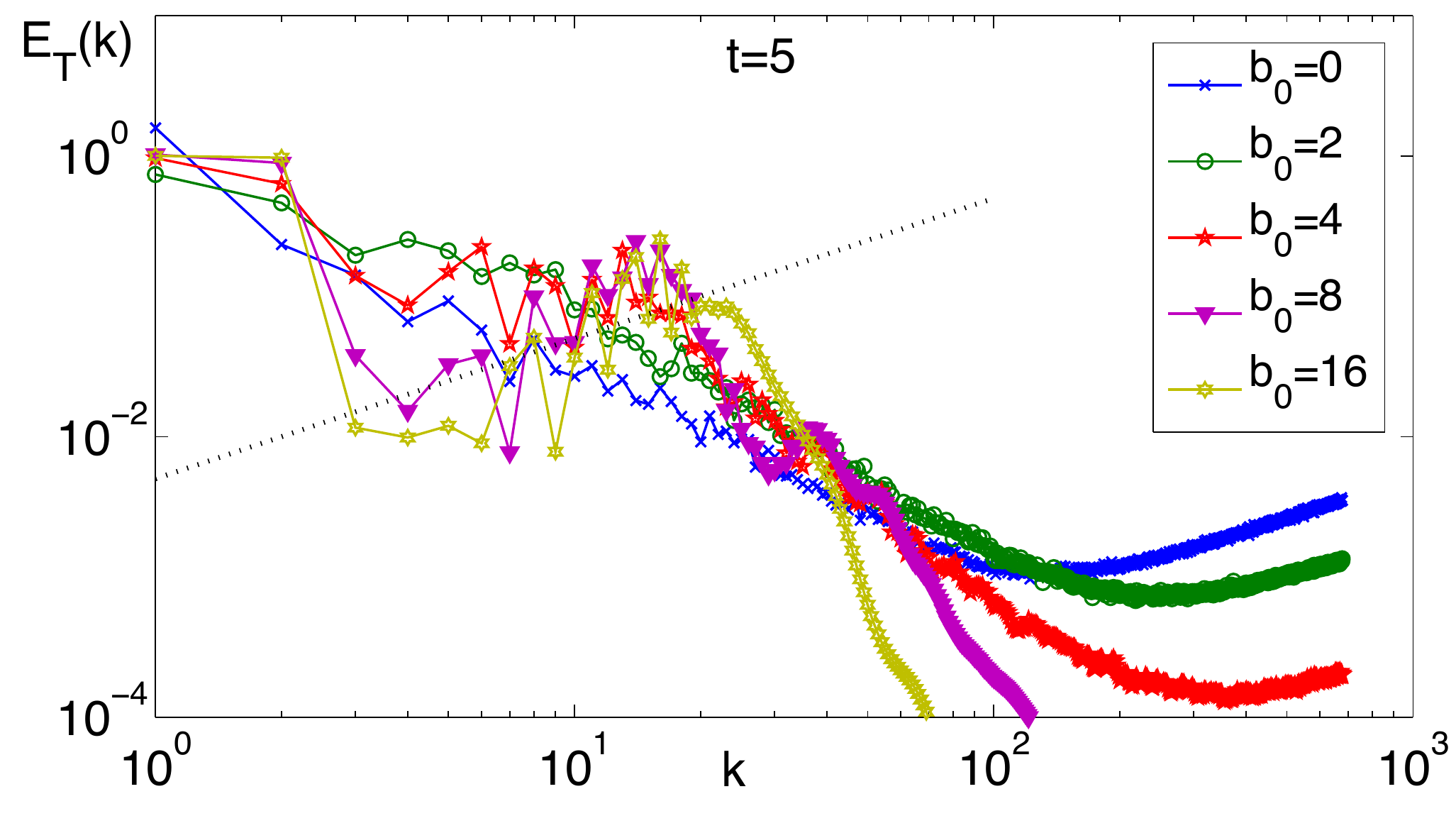}
\includegraphics[width=0.49\textwidth]{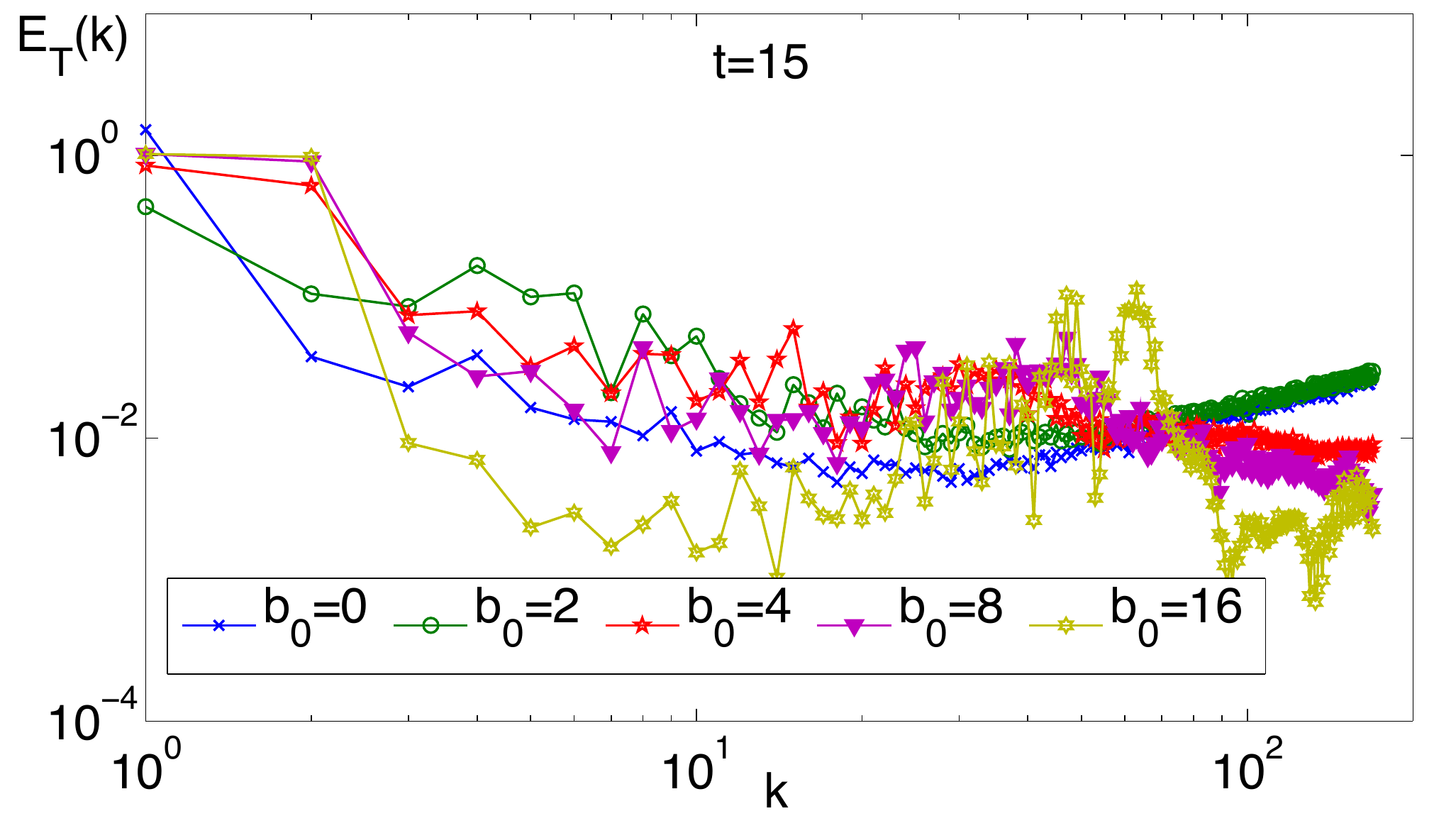}  \includegraphics[width=0.49\textwidth]{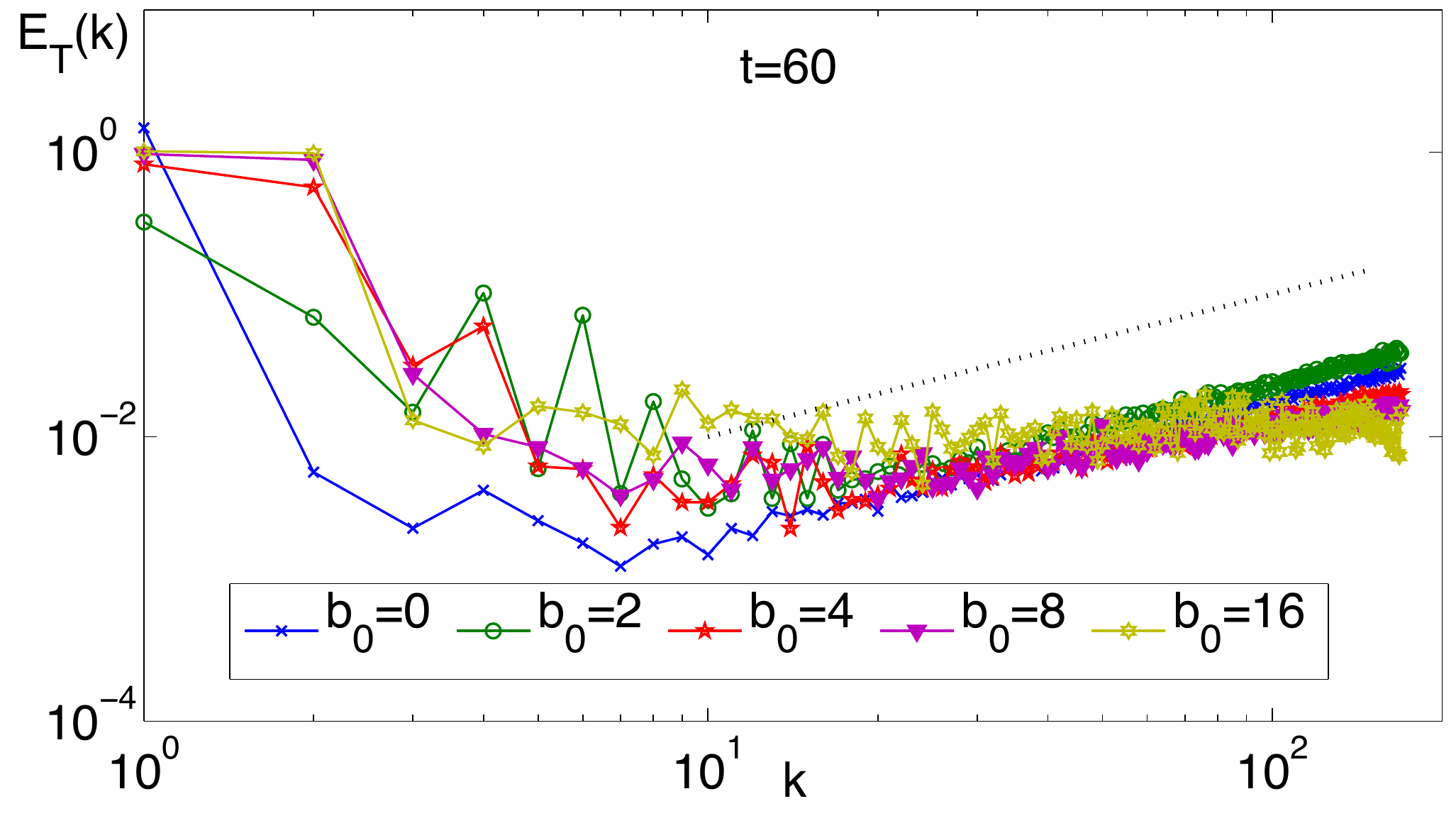}
\caption{Total energy spectra with $b_0 =0$ (x), $b_0 =2$ (circle), $b_0 =4$ ($\ast$), $b_0 =8$ (triangle) and $b_{0}=16$ (star) at $t=1.2$ and $t=5$ (top, with a grid of $2048^2$ points, Runs 3 and 4), and $t=15$ and $t=60$  (bottom, grid of $512^2$ points, Runs 2). The dashed lines correspond to a $k^1$ scaling. Thermalization may occur for very long times, but the spectral index at intermediate scales/times for higher $b_0$ is far from the predicted value, with possibly a partial thermalization at these intermediate scales (see top, right).
} \label{Fig:SpAlfven}  \end{figure*} 

It may appear somewhat surprising that a different conclusion is reached with the present data from what is argued in \cite{FRISCH:1983p3300}  on the basis of a quasi-regular behavior of MHD in two-dimensions in the presence of a large-scale magnetic field: indeed, such a field retards the nonlinear dynamics and can be seen as a bath of weakly interacting Alfv\'en waves with a spectrum that can be derived analytically in the case of a strong enough $b_0$, using weak turbulence theory \cite{Galtier:2000p3795}. 
However, it is well-known that the weak turbulence approach is non-uniform in scale: the small parameter of the problem, $R_{WT}$, should be evaluated taking into account that the eddy turn-over time gets smaller as smaller scales are reached, whereas the length scale of the imposed field remains infinite,
 by construction. Working out this relation in the case of the IK spectrum, $E(k)\sim (\epsilon b_0)^{1/2} k^{-3/2}$ leads to a scale $\ell_B\sim 1/b_0$ with $R_{WT}(\ell_B)=1$, scale beyond which a classical small scale turbulent spectrum will develop; it should be noted, however, that the resolution of the computation must be such that $\ell_B$ is reachable accurately 
($2\ell_B k_{\rm max}> 1$) in order to observe this phenomenon.

We now examine  numerically the spectral relaxation to equilibrium; several times are displayed in Fig.\ref{Fig:SpAlfven}, with dashed lines corresponding to a $k^{+1}$ scaling.
 We first observe that, at early times, the spectra coincide at large scales, but that small-scale thermalization is delayed for stronger imposed fields.

\begin{figure}
    \includegraphics[width=0.48\textwidth]{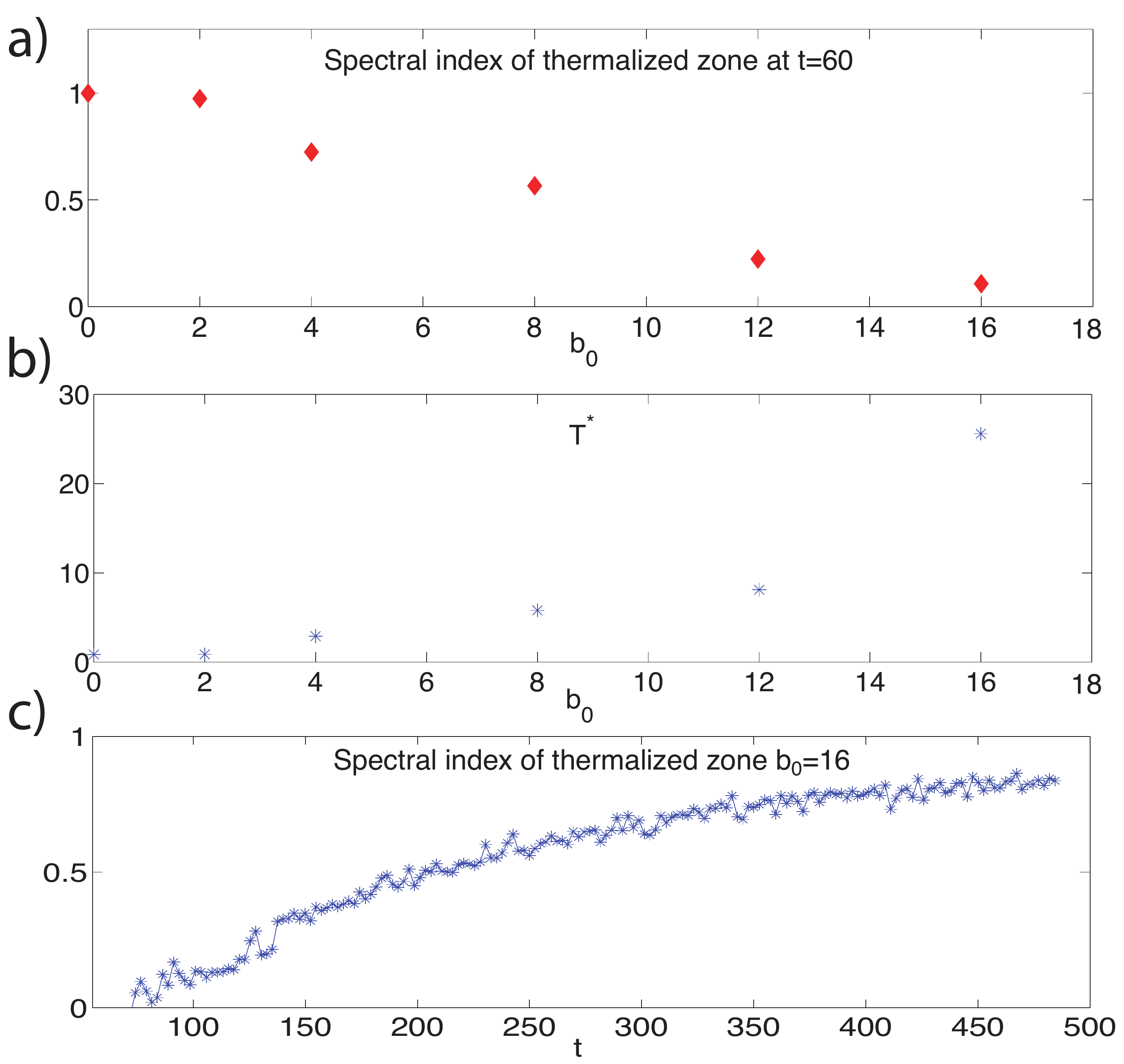} 
   \caption{Variations with imposed mean field $b_0$.
    a) Spectral index $\alpha=-m$ (see Eq.\eqref{Eq:delta0} with $\delta(t)\equiv0$) of the thermalized part ($k\ge20$) of the total energy spectra at $t=60$ (see Fig.\ref{Fig:SpAlfven}) for several $b_0$.  Orszag-Tang vortex, resolution of $512^2$ grid points, Runs 2 (see Table \ref{Table:CompNLS-NGLR}).
   b) Time $T^{\ast}$ as a function of $b_0$; $T^{\ast}$ is defined as the time at which the grid size is reached by the computation, as measured by the logarithmic decrement $\delta$.
c) Temporal evolution of spectral index for $k\ge20$ with an imposed uniform field $b_0=16$ at resolution $256^2$ (Run 1); note the slow evolution toward $\alpha=1$ as expected at equilibrium.
  }\label{Fig:Loidechelle} \end{figure}

 At intermediate times, there is a domain of wavenumbers in which the dynamics differs quite substantially according to the value of ${\bf b}_0$. It appears that there is now a relaxation to equilibrium at intermediate wavenumbers (the $k^1$ scaling law seems to be followed, see the figure) before a sharp plunge in the spectrum due to the fact that small scales are not reached yet because of the substantial slowing down of the dynamics. This is particularly striking at $t\sim 5$ for which the run with the largest value of the imposed mean field already presents a partial thermalization at intermediate wavenumbers, as if the dynamics was indeed seeing a truncation at wavenumbers much lower than the actual $k_{max}$ of the run, whereas for ${\bf b}_0=4$, energy continues to flow to smaller scales. At a later time ($t\sim 15$), the ${\bf b}_0=4$ case now may be seeing a pseudo-truncation in scale and partially thermalizes at intermediate scales, but it still cannot reach the smallest scales numerically available to the run. 
 At the final time of the computation, $t\sim 60$,  the three runs start to display thermalization in a broader range of scales, although the evolution of the gravest mode is quite retarded again compared to the weak field case. 
However note that the exponent of the thermalized zone clearly depends on ${\bf b}_0$, (see Fig.\ref{Fig:Loidechelle}.a); for ${\bf b}_0=16$, the exponent is close to $0$ indicative of a partial one-dimensional thermalization.

Figure.\ref{Fig:Loidechelle}.b displays the time $T_{\ast}$ at which the smallest effective excited scale in the flow reaches the grid size ($\delta (T_{\ast})=2/k_{max}$). Observe that $T_{\ast}$ is an increasing function of $b_0$ and no sign of saturation of this slowing-down is observed yet, although one might want to test higher values of $b_0$ as well. Including all points, a quadratic variation of $T_{\ast}$ with $b_0$ is plausible.
Note however, that the close to zero value of the spectral index for $b_0=16$ (see Fig.\ref{Fig:Loidechelle}.a) is in apparently contradiction with the expected $k^{+1}$ power-law of equipartition of energy in $2D$. It requires substantially longer times of integration to check whether the convergence toward the predicted statistical equilibria is particularly slow or whether another solution is obtained (as in the Fermi-Pasta-Ulam-Tsingou problem \cite{Fermi:1955p3871}). In order to investigate this point, we performed an integration until $t=475$ at moderate resolution. The spectral index for this run is displayed in Fig.\ref{Fig:Loidechelle}.c and it is found to asymptotically approach to $1$, the predicted value of the $2D$ absolute equilibrium.

\section{Structures} \label{s:struct}

{The structures that develop in the flow we study here have been examined in detail in \cite{FRISCH:1983p3300} at early time. Here, we pursue this study at higher resolution and examine how these structures change with ${\bf b_0}$ and with time.
The structures are shown in Fig. \ref{Fig:Struc} using grids of $2048^2$ and $4096^2$ points 
(note that the color bars are described in the caption). 
 \begin{figure} 
 \includegraphics[width=.225\textwidth]{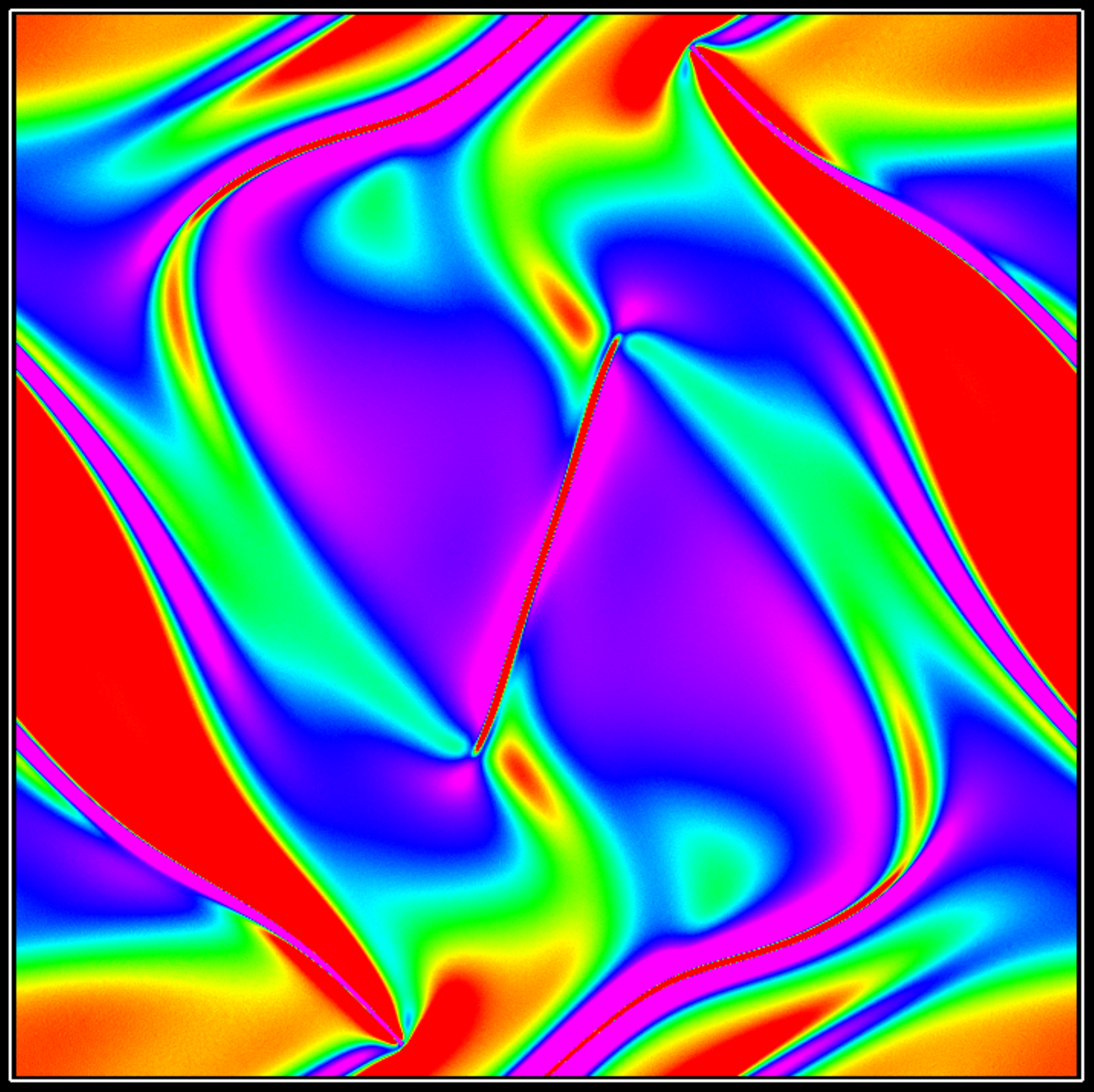}  
 \includegraphics[width=.225\textwidth]{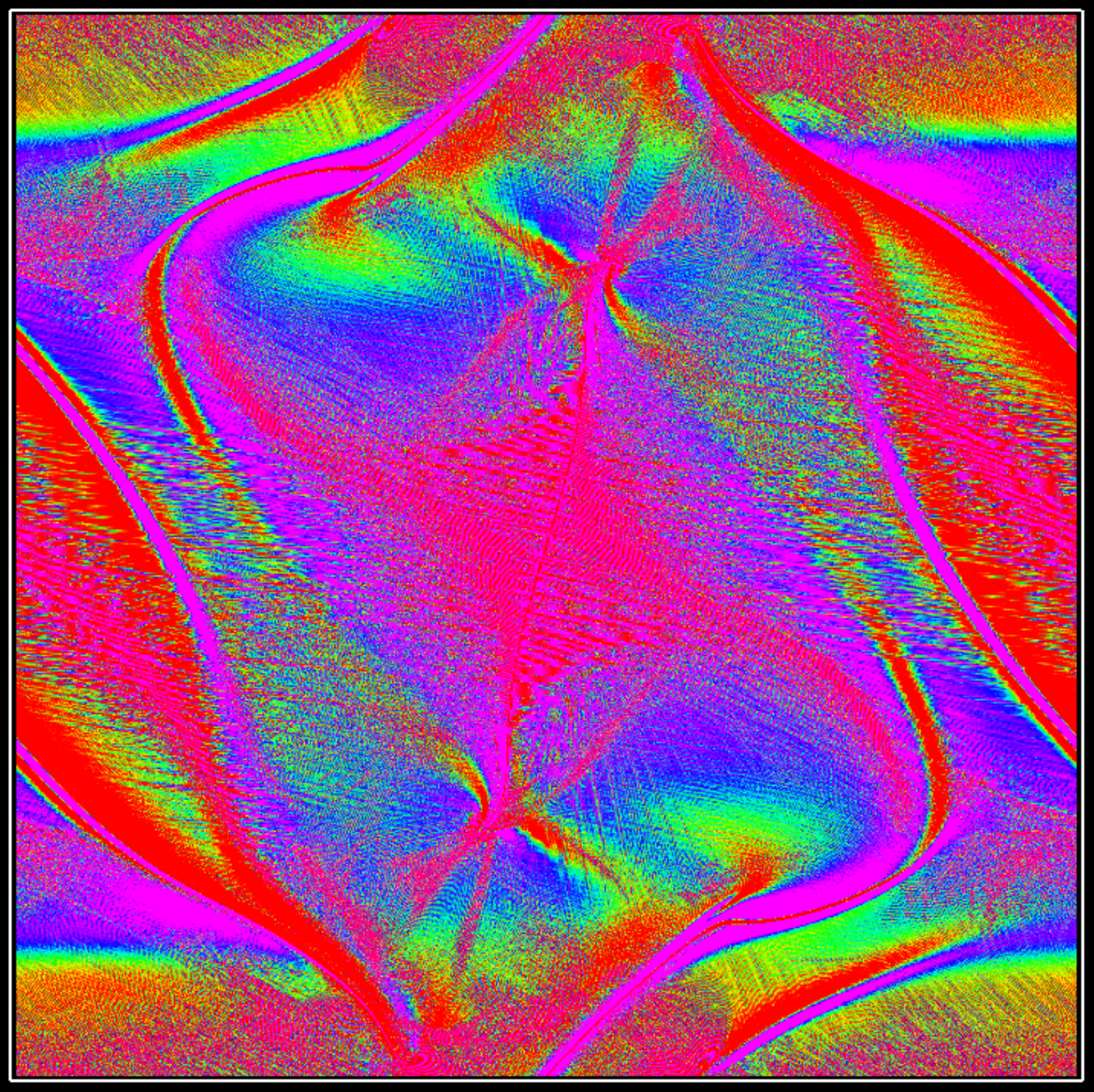} 
\includegraphics[width=.225\textwidth]{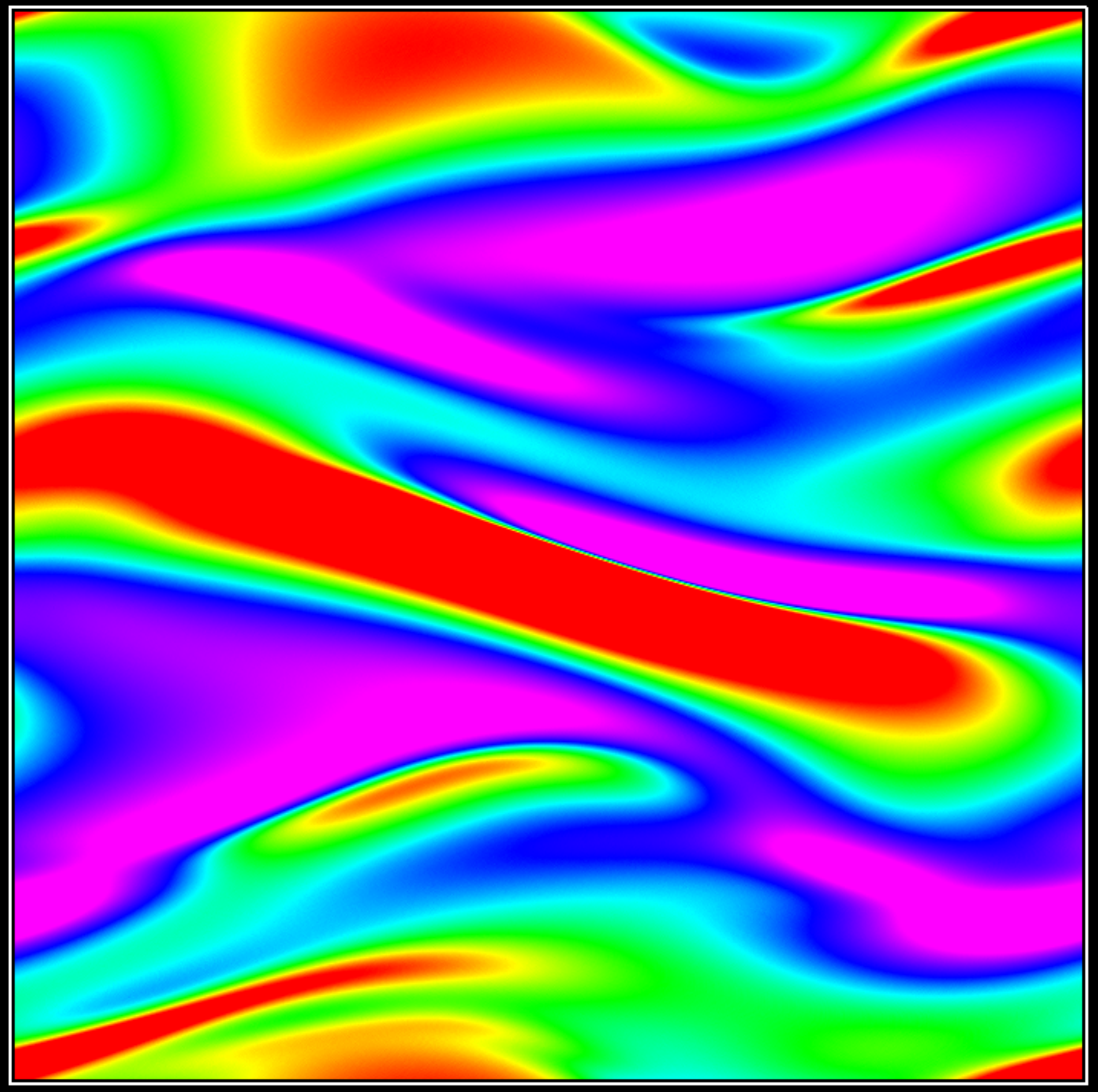} 
\includegraphics[width=.225\textwidth]{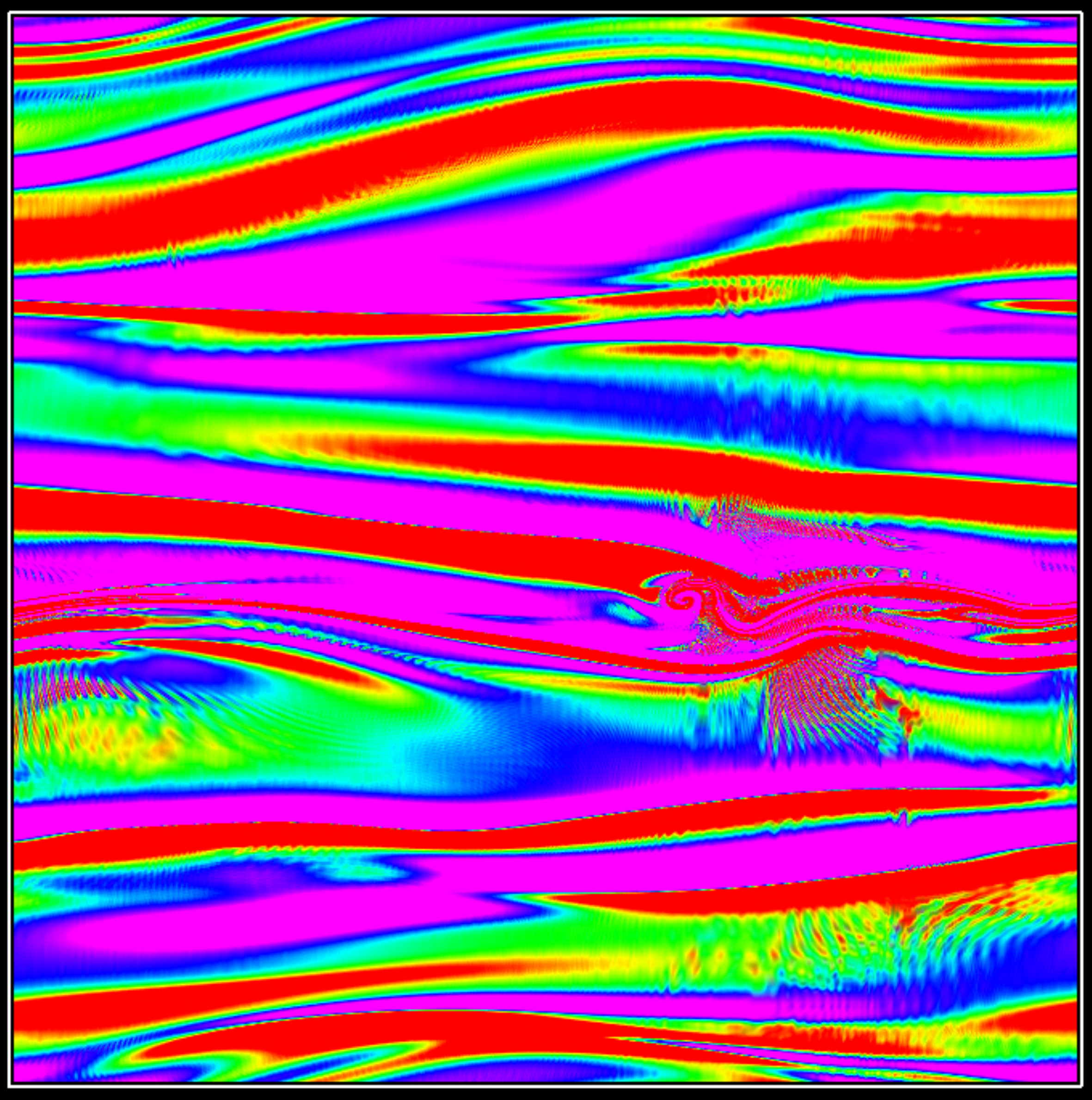}                       
 \caption{The color bars (not shown) are such that red and magenta represent extreme values, green and blue moderate to low values of the fields:
 stream function and  vorticity vary from -4 (red) to 4 (purple), with in the former case most of the points at values close to zero, whereas the vorticity in this range has all values with close to equal probability;
the magnetic potential varies from -3, red, to 3, purple, and  the current from -6, red, to 6, purple.
  {\it Top:} Current for $b_0=0$, $t=0.9$ (left) and $t\sim 1.1$ (right). Resolution of $4096^2$ grid points.
  {\it Bottom:} Current for $b_0=4$, $t=1.1$ (left) and $t\sim 3.7$ (right). Resolution of $2048^2$ grid points.
  In the presence of even a moderate uniform field ($b_0/b_{rms}\sim 1.6$ here), the onset of thermalization that destroys structures is delayed and the flows align themselves with $b_0$.}
         \label{Fig:Struc} 
\end{figure}    
}
At early times, the flows are made up of quasi-singular structures which persist until $t\lesssim1$ for $b_0=0$ (Figs. \ref{Fig:Struc} top) and for longer times  when the magnitude of the external field is increased (see Figs. \ref{Fig:Struc} bottom, obtained for the case $b_0=4$). Once the thickness of these structures reach the grid size, noise steps in but the structures continue to evolve until the system is able to reach complete thermalization.

In the presence of an imposed field, when strong enough, structures align themselves in its direction. Furthermore, the appearance of noise is clearly delayed; at a given time, it is not yet visible as the amplitude of $b_0$ increases (contrast the current at $t=1.1$ for $b_0=0$ (top-right) and $b_0=4$ (bottom-left)). Finally, note that in all cases with strong imposed field, the small-scale structures display an intense folding and piling-up of sheets of opposite signs.

\section{Conclusions} \label{s:conclu}
 
We have investigated in this paper the dynamics of an ideal two-dimensional fluid in the MHD limit, in the presence or not of a uniform magnetic field ${\bf b}_0$,  and we have shown the link to the dissipative driven (DD) case. For ${\bf b}_0\equiv 0$ and at intermediate times and intermediate scales, a behavior observed in the DD case obtains, namely that the energy spectrum is that proposed by Iroshnikov and Kraichnan, with a $k^{-3/2}$ power law   \cite{Iroshnikov:1964p3299, Kraichnan:1965p2693} 
for the Orszag-Tang vortex and for random flows, with in both cases initial conditions centered in the large scales.
Furthermore, as already found in \cite{FRISCH:1983p3300} but for shorter evolution times and lower resolutions, the formation of small scales is inhibited when ${\bf b}_0\not= 0$. 
 
Note that small-scale initial data was also studied in \cite{Krstulovic:2010Thesis} where it was shown that the system reaches equilibrium by an eddy-noise mechanism. No inverse cascade was observed as mechanism of thermalization. In the case of MHD an equipartition of magnetic potential leading to a $k^3$ scaling-law for the magnetic energy spectrum and $k^1$  equipartition for the kinetic energy spectrum was obtained at large times.

Extension of this work to the three-dimensional case in MHD may be of use for at least three reasons:

\noindent (i) It has been shown in \cite{lee2} that, for initial conditions that are identical from the point of view of the statistics (same energy, same velocity field, same equipartition between kinetic and magnetic energy at $t=0$, same initial conditions centered in the  large-scales, same total magnetic helicity, and with total velocity-magnetic field correlation between 0 and 4\%, in normalized value), three different energy spectra could emerge in the absence of imposed uniform magnetic field and forcing (decay case with non-zero viscosity and unit magnetic Prandtl number) when considering three different initial conditions for the induction. Would they be observed as well at intermediate times- intermediate scales in the ideal case? It is plausible to think so, since a Kolmogorov spectrum is observed in the ideal 3D case for neutral fluids, and we observe the Iroshnikov-Kraichnan spectrum in the present work, but it would be of interest to verify the lack of universality in MHD in the ideal case as well.

\noindent (ii) Would a fast-decreasing spectrum at the end of the inertial range and before the thermalized range obtain in three-dimensional MHD, as it does in the 3D neutral case? It is argued in this paper that the lack of such a range is probably due to an insufficient number of thermalized modes because the total number of modes in two dimensions is not very large, compared to the three-dimensional case at the same linear resolution; thus, an effective eddy viscosity does not obtain here, and such a 3D computation would provide a test of this idea.

\noindent (iii) Finally, the case of magnetic helicity $H_M=<{\bf a} \cdot {\bf b}>$, an invariant in ideal MHD in three dimensions, deserves a separate study. Indeed, it was shown in \cite{PFL} that, performing a standard phenomenology \`a la Kolmogorov on $H_M$, the spectrum becomes $H_M(k)\sim k^{-7/3}$ but recent studies have shown that different spectra obtain \cite{mueller, mininni2, malapaka} with $H_M(k)\sim k^{-3}$ or steeper. The origin of this discrepancy is not completely understood; it could be related to a detailed equipartition between kinetic and magnetic modes in the energy and helicity parts of their spectral correlation functions. Again, an ideal study may help unravel the mechanisms at play in the dynamical evolution of MHD turbulence.

\vskip0.5truein

\begin{acknowledgments}
Computer time was provided by IDRIS.
The National Center for Atmospheric Research is sponsored by the National Science Foundation. 
\end{acknowledgments}

\appendix

\begin{figure} 
  \includegraphics[width=0.49\textwidth]{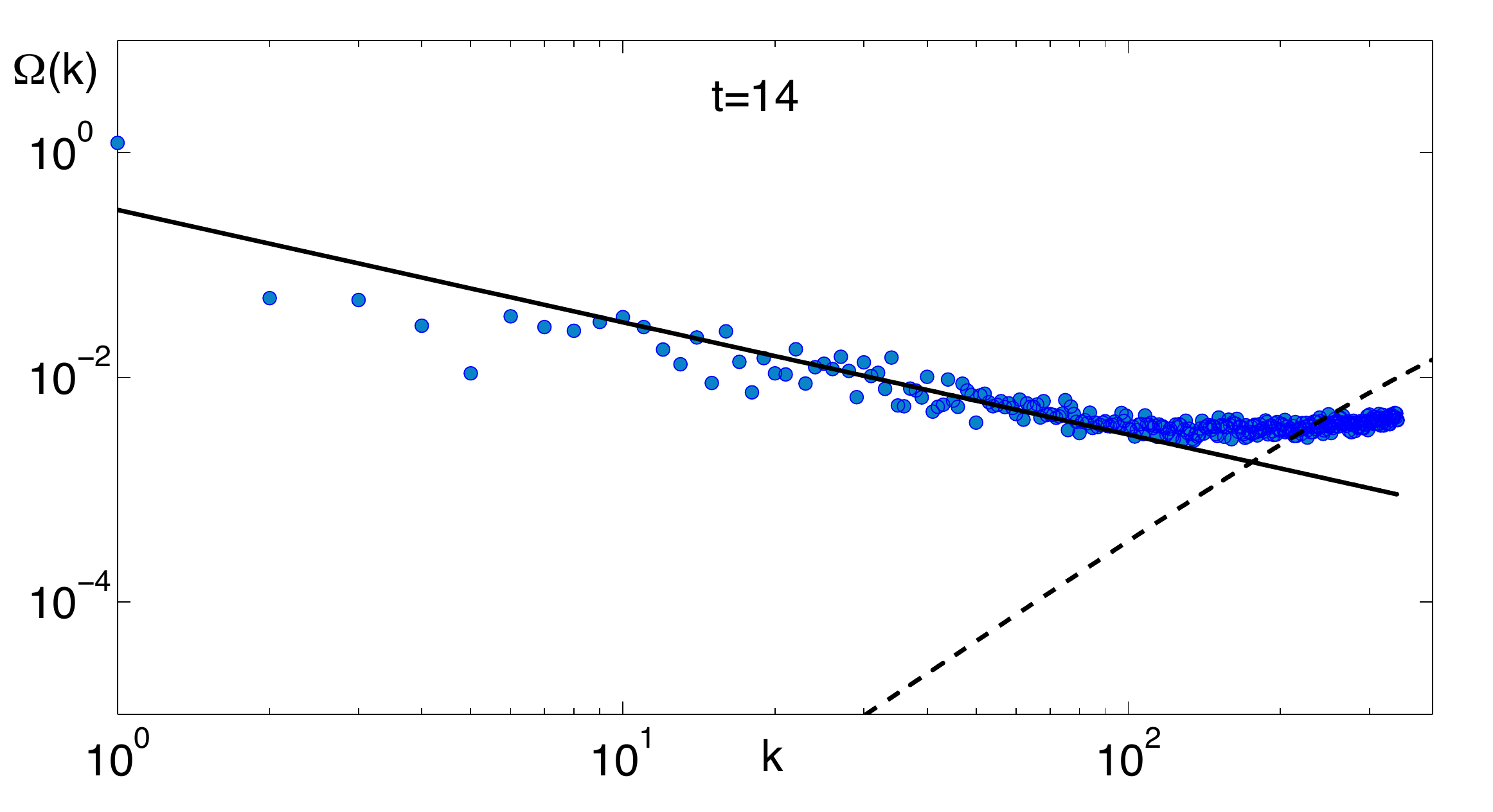}  \includegraphics[width=0.49\textwidth]{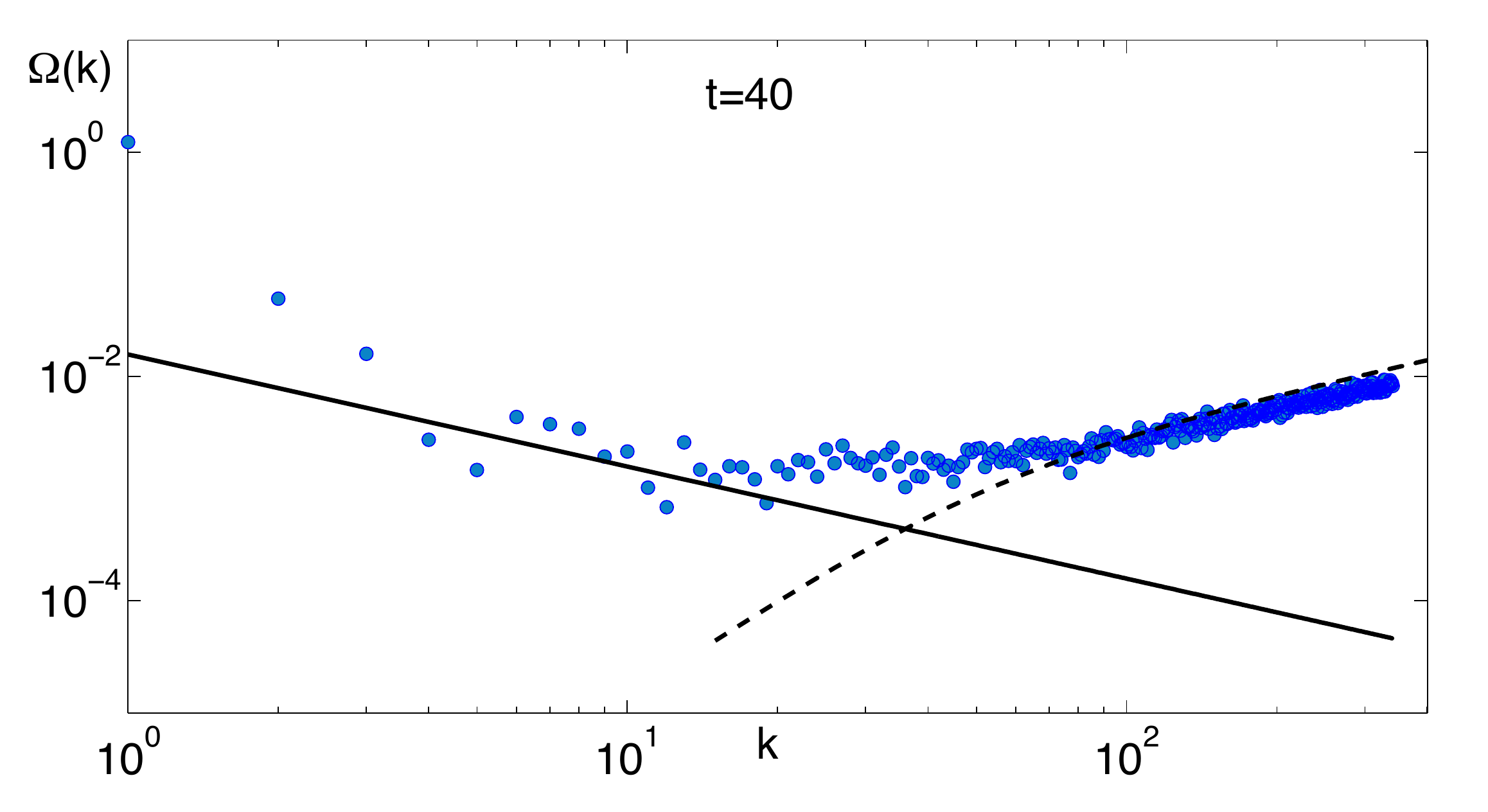}
  \caption{Temporal evolution of enstrophy spectra $k^2E(k)$ for 2D Euler, at $t=14$ (top) and $t=40$ (bottom). Solid 
  and dashed lines respectively indicate the enstrophy cascade ($k^{-1}$-scaling) and thermalized enstrophy at small scales. Again note, as in the 2D MHD case, the absence of a pseudo-dissipative range before equipartition sets in.}
\label{Fig:SpecEnstrophy} 
\end{figure}

\section{The fluid case in 2D \label{Sec:Euler}}

We take here as initial conditions, ${\bf b}\equiv 0$, and for the stream function we have:

\begin{equation}
\psi(x,y)=\frac{1}{k_a}\sin{k_ax}\sin{k_ay}+\frac{2}{k_b}\cos{k_bx},\label{Eq:IniCond}
\end{equation}
with the parameters set to $k_a$=1 and $k_b$=2.
The temporal evolution of the enstrophy spectra is displayed in Fig. \ref{Fig:SpecEnstrophy}. As in 3D, a clear scale separation also appears: there is a progressive thermalization starting from the smallest scales, with the energy (vs. enstrophy) cascading to the larger (vs. smaller) scales.
Note that the thermalized enstrophy increases from zero at early times to an amount of the order of the total enstrophy available in the system. Defining $k_{th}$ as previously, and using the values of $\Omega_{\rm  th}$ and $E_{\rm  th}$, we can compute the parameters (Lagrange multipliers) $\alpha$ and $\beta$ from: 
 \begin{equation}
E(k)=\frac{2\pi k}{\alpha+\beta k^2} \ , \  \Omega(k)= k E(k) \ .
 \label{Eq:EqAbsEnstro} \end{equation}
 
 These Kraichnan absolute equilibria are displayed as solid lines at small scale in Fig. \ref{Fig:SpecEnstrophy} (notice that they curve down at the larger scale of the thermalized zone for enstrophy); these solutions correspond to the $k^3$-scaling of the high enstrophy containing absolute equilibria. 
 
 The good agreement shows that the  evaluation of $k_{th}$ and the energy and enstrophy at the wavenumber describe reasonably correctly the temporal behavior of the flow (in the latter case, as $\eta_{th}^{2/3}k^{-1}$). The enstrophy dissipation rate  $\eta_{\rm th}$ can be estimated, defining it as the time derivative of $\Omega_{\rm  th}$. The respective spectra are also displayed as solid lines at large scales in Fig. \ref{Fig:SpecEnstrophy}; observe that in the inertial zone both scaling law and prefactor are in good agreement with  $\Omega(k)\sim\eta_{\rm th}^{2/3}k^{-1}$.
 
The relaxation dynamics of two-dimensional Euler turbulence is, \emph{mutatis mutandis}, similar to the three-dimensional Euler case: a direct cascade of enstrophy (energy in 3D) followed by an equilibration of enstrophy (energy in 3D) at small scale. It can be shown \cite{Krstulovic:2010Thesis} that the dissipation wavenumber estimated from the equilibria is of the order of the maximum wavenumber, thereby explaining in this case the absence of dissipation range as a buffer zone between  the inertial range and the thermalized range, contrary to the 2D MHD case.


\end{document}